\documentclass[aip,jcp,amsmath,amssymb,reprint]{revtex4-1}

\newcommand{\Nm}{N_{\mathrm{m}}}

\newcommand{\rc}{r_{\mathrm{c}}}
\newcommand{\Tc}{T_{\mathrm{c}}}
\newcommand{\kTc}{k_{\mathrm{B}}T_{\mathrm{c}}}
\newcommand{\kBT}{k_{\mathrm{B}}T}
\newcommand{\mc}{m_{\mathrm{c}}}

\newcommand{\trd}{\tilde r_{\mathrm{d}}}
\newcommand{\wwd}{w_{\mathrm{d}}}
\newcommand{\kapnd}{\kappa_{\text{nd}}^{-1}}
\newcommand{\alphaMB}{\alpha_{\rm{MB}}}
\newcommand{\talphaMB}{\tilde\alpha_{\rm{MB}}}

\newcommand{\vc}[1]{\mathbf{ #1}}
\newcommand{\p}[1]{\partial{#1}}

\usepackage{graphicx,dcolumn,bm}
\usepackage{sidecap}
\usepackage{booktabs,array}
\usepackage{siunitx}
\usepackage{subcaption}
\usepackage{hyperref}
\hypersetup{colorlinks,}

\begin{document}
\title{Invariance of experimental observables with respect to coarse-graining in standard and many-body dissipative particle dynamics}

\author{Peter Vanya}
\email{peter.vanya@gmail.com}
\affiliation{Department of Materials Science \& Metallurgy, University of Cambridge, 27 Charles Babbage Road, Cambridge CB3 0FS, United Kingdom}

\author{Jonathan Sharman}
\affiliation{Johnson Matthey Technology Centre, Blounts Court Road, Sonning Common, Reading RG4 9NH, United Kingdom}

\author{James A. Elliott}
\email{jae1001@cam.ac.uk}
\affiliation{Department of Materials Science \& Metallurgy, University of Cambridge, 27 Charles Babbage Road, Cambridge CB3 0FS, United Kingdom}

\date{\today}

\begin{abstract}
Dissipative particle dynamics (DPD) is a well-established mesoscale simulation method. However, there have been long-standing ambiguities regarding the dependence of its (purely repulsive) force field parameter on temperature as well as the variation of the resulting experimental observables, such as diffusivity or surface tension, with coarse-graining (CG) degree. Here, we revisit the role of the CG degree and rederive the temperature dependence in standard DPD simulations. Consequently, we derive a scaling of the input variables that renders the system properties invariant with respect to CG degree, and illustrate the versatility of the method by computing the surface tensions of binary solvent mixtures. We then extend this procedure to many-body dissipative particle dynamics (MDPD) and, by computing surface tensions of the same mixtures at a range of CG degrees, demonstrate that this newer method, which has not been widely applied so far, is also capable of simulating complex fluids of practical interest.
\end{abstract}

\maketitle

\section{Introduction}
Coarse-grained molecular dynamics (MD) contains, in addition to the usual force field- or thermostat-related parameters associated with atomistic MD, another parameter: coarse-graining (CG) degree, which provides the freedom to trade off between simulation speed and spatial or temporal resolution. As CG degree is a theoretical construct without experimental substance, the physical properties of a simulated system must be the same, or at least as similar as possible, at different scales.

Materials simulations are usually performed using reduced units in order to avoid extremely small or large numbers and prevent the duplication of thermodynamically equivalent states. Some CG simulation methods, such as dissipative particle dynamics (DPD) widely used for soft matter, work in units where the length scale is defined from the reduced density and CG degree, both of which one is free to choose.\cite{Groot_JCP_1997} As a result, the conversion from real to reduced units while preserving the physical properties has been rather ambiguous and the comparison of the same physical systems simulated at different CG degrees nearly impossible. The purpose of the present paper is to address this challenge through a consistent scaling approach.

The standard version of DPD has been successfully applied to a wide range of soft matter systems in the past two decades.\cite{Espanol_JCP_2017} On top of it, a many-body dissipative particle dynamics (MDPD) method has been proposed by adding a density-dependent term into the force field.\cite{Pagonabarraga_JCP_2001,Trofimov_JCP_2002,Warren_PRE_2003} This version is capable of simulating non-ideal fluids and free surfaces, and hence covers a much wider range of systems of practical interest.\cite{Ghoufi_EPJ_2013}

Unfortunately, no general protocol for deriving the MDPD interaction parameters for real materials has been proposed so far. In case of standard DPD, the sole interaction parameter $a$ is obtained by matching the compressibility to an equation of state (EOS) of a pure liquid, and cross-interaction parameters for mixtures are based on some mean-field approximation, such as Flory-Huggins theory. However, in case of MDPD, the choice has so far been \emph{ad hoc}.~\cite{Ghoufi_PRE_2011,Ghoufi_PRE_2010,Ghoufi_PRE_2010,Ghoufi_JCTC_2012}

Furthermore, there has been much discussion about how the DPD interaction parameter should scale with CG degree. Groot \& Rabone\cite{Groot_BiophysJ_2001} originally suggested a linear dependence, but this was refuted independently by Maiti \& McGrother\cite{Maiti_JCP_2004} and F\"uchslin \emph{et al.}\cite{Fuchslin_JCP_2009} Maiti \& McGrother also proposed linear scaling for the $\chi$-parameter with the aim of reproducing the experimental surface tensions. However, we have found inconsistencies in their reasoning. In case of MDPD, to our knowledge there have been no predictions of the surface tension for real mixtures and no discussion of the scaling.

The aim of this paper is to present a general protocol to determine the interaction parameters as a function of not only material properties, such as compressibility or surface tension, but also the coarse-graining degree and temperature. To achieve this, we first need to discuss the choice and role of the reduced units. We revisit and restate the derivation presented by F\"uchslin \emph{et al.}, and correct the temperature-dependence of the interaction parameter originally proposed by Groot \& Warren. More importantly, we explain the derivation and the reasoning behind the scaling of the Flory-Huggins $\chi$-parameter, a key variable determining the mixing properties of soft matter. Consequently, we are able to demonstrate the improved predictive accuracy of these methods across a range of CG degrees.

The paper is structured as follows. In Section~\ref{sec:dpd}, we revisit F\"uchslin \emph{et al.}'s arguments for the scaling in DPD and the derivation of the interaction parameters as a function of temperature. In Section~\ref{sec:dpd_rescale_expt}, we present a method to achieve the scale invariance of experimental observables. Section~\ref{sec:mdpd} extends the parametrisation protocol to MDPD and establishes that this method is capable of predicting the surface tension of binary solvent mixtures.

\section{Summary of reduced units and parameterisation in DPD}
\label{sec:dpd}
We denote all variables in reduced units by a diacritical tilde, following the convention set by F\"uchslin \emph{et al.}\cite{Fuchslin_JCP_2009} Defining a set of reduced units $\rc=1, \mc=1, \kTc=1$, where $\Tc$ is a user-selected reference temperature, the conversion is, for example, $\tilde r = r/\rc$, $\tilde m=1=m/\mc$, since all the DPD bead masses are assumed to be the same, and $\widetilde{\kBT} = \kBT/\kTc$, where $\kTc$ is the energy scale based on the chosen temperature. Note that for constant-temperature simulations it is possible to set $\kBT = \kTc$, in which case the reduced temperature $\widetilde{\kBT}=1$. However, in general case it is useful to distinguish between the energy scale $\kTc$ and temperature $\kBT$.

We first briefly describe the DPD force field. Its hallmark is the linear dependence of force on distance:
\begin{equation}
\vc{\tilde F}^{\rm C}(\vc{\tilde r}) = 
\begin{cases}
\tilde a w(\tilde r) \vc{\hat r}, & \tilde r \leq 1 ,\\
0, & \tilde r > 1,
\end{cases}
\end{equation}
where:
\begin{equation}
w(\tilde r) = \begin{cases}
1 - \tilde r, & \tilde r \leq 1,\\
0,            & \tilde r > 1
\end{cases}
\end{equation}
is the weight factor, $\vc{\tilde r} = \vc{\tilde r}_i - \vc{\tilde r}_j$ interparticle distance, $\tilde r=|\vc{\tilde r}|$ vector magnitude, $\vc{\hat r} = \vc{\tilde r}/\tilde r$, and $\tilde a$ a parameter representing the interaction strength. In the simulation, the conservative force is complemented by a Langevin-type thermostat with dissipative and random forces:
\begin{align}
\vc{\tilde F}^{\rm D}(\vc{\tilde r}) &= -\tilde\gamma w(\tilde r)^2 (\vc{\hat r}\cdot \vc{\tilde v}) \vc{\hat r},
\label{eq:fd}\\
\vc{\tilde F}^{\rm R}(\vc{\tilde r}) &= \sqrt{2\tilde\gamma\widetilde{\kBT}} w(\tilde r) \frac{\theta}{\sqrt{\Delta \tilde t}} \, \vc{\hat r},
\label{eq:fr}
\end{align}
where $\tilde\gamma$ is the friction parameter, $\vc{\tilde v} =\vc {\tilde v}_i - \vc{\tilde v}_j$ the relative particle velocity, $\theta$ a Gaussian random number with zero mean and unit variance, and $\Delta \tilde t$ a simulation step, i.e. $\theta/\sqrt{\Delta \tilde t}$ is a Wiener process. The purpose of the term $\sqrt{\Delta \tilde t}$ in the denominator of eq.~\eqref{eq:fr} is to enforce diffusion independent of time step in numerical simulations with finite precision. This point is thoroughly clarified in Ref.\cite{Groot_JCP_1997}

In the case of a single-component fluid, the DPD field is sufficiently simple that its EOS can be easily reverse-engineered, as done by Groot \& Warren (GW):\cite{Groot_JCP_1997}
\begin{equation}
\tilde p = \tilde\rho \widetilde{\kBT} + \tilde\alpha \tilde a \tilde\rho^2,
\end{equation}
where $\tilde\rho$ is the number density and $\tilde\alpha$ is a fitting constant, which was shown to be approximately 0.1 for $\tilde\rho>3$. 

To derive $\tilde a$, these authors matched the EOS to the isothermal compressibility $\kappa$:
\begin{equation}
\kappa^{-1} = \rho \bigg( \frac{\p p}{\p\rho}\bigg)_T,
\end{equation}
which leads to an interaction parameter in reduced units $\tilde a = 25\widetilde{\kBT}$ at a reduced density $\tilde\rho=3$ (shown on the last line of Section IV of their paper\cite{Groot_JCP_1997}).

% delete
%To derive $\tilde a$, these authors matched the EOS to the isothermal compressibility $\kappa$. From definition,
%\begin{equation}
%\kappa^{-1} = \rho \bigg( \frac{\p p}{\p\rho}\bigg)_T,
%\end{equation}
%and, in reduced units,
%\begin{equation}
%\tilde\kappa^{-1} = \tilde\rho \bigg( \frac{\p{\tilde p}}{\p{\tilde\rho}}\bigg)_{\tilde T} = \tilde\rho \widetilde{\kBT} + 
%2\tilde\alpha\tilde a \tilde\rho^2.
%\end{equation}
%Considering water with compressibility $\kappa\approx 4.5\times 10^{-10}$~Pa$^{-1}$, which can be non-dimensionalised to $\kapnd = 1/(\kappa  n \kBT)$, where $T$ is absolute temperature and $n$ is molecular number density, and making the choice that one DPD particle (\emph{bead}) contains one molecule, the interaction parameter $\tilde a$ derived by GW for room temperature, when $\kapnd\approx 16$ (below eq.~(16) in their paper\cite{Groot_JCP_1997}), is:
%\begin{equation}
%\tilde a = \frac{\kapnd - 1}{2\tilde\alpha\tilde\rho} \widetilde{\kBT} 
%= 25\, \widetilde{\kBT}.
%\end{equation}
%Note that $a$ has the dimension of force, i.e. $\kTc/\rc$, but $\rc$ in the denominator is set to 1.
% end delete

To bridge the simulation method with real materials, Groot \& Rabone defined the length scale (and interaction cutoff at the same time) $\rc$ as follows:\cite{Groot_BiophysJ_2001}
\begin{equation}
\rc = (\tilde\rho\Nm V_0)^{1/3},\quad\text{\emph{ i.e. }}\quad \rc \sim \Nm^{1/3},
\end{equation}
where $\Nm$ is the CG defined as the number of molecules in one DPD bead and $V_0$ is the volume of a single water molecule. These authors consequently derived that the parameter $\tilde a$ should scale linearly: $\tilde a(\Nm) = \Nm \tilde a(1)$. However, their reasoning was refuted by F\"uchslin \emph{et al.}, who showed that, in real units, the scaling is a power law: $a(\Nm) = \Nm^{2/3} a(1)$. More importantly, F\"uchslin \emph{et al.} showed that in reduced units the interaction parameter does \emph{not} scale: $\tilde a(\Nm) = 25$ for any choice of $\Nm$. This is a very useful feature, as one is now free to simulate a pure liquid at any CG degree without worrying about inducing undesirable simulation artefacts such as freezing, which can happen for $\tilde a>200$.\cite{Trofimov_PHD_2003}

\subsection{Scaling with coarse-graining degree}
\label{sec:dpd_scaling}
Here, we rederive the scaling with respect to the CG degree $\Nm$. The purpose of this analysis is to provide a simple and robust framework to understand the scaling of any variable of interest. We reproduce the derivation due to F\"uchslin \emph{et al.}\cite{Fuchslin_JCP_2009} with simple arguments of dimensionality. Thus we will be able to track the scaling of separate variables, which would otherwise become overly complicated since the length scale $\rc$ depends on $\Nm$. 

As a first step, we convert the density from reduced to real units. Knowing that the density of unscaled liquid with $\Nm=1$ is the same as the number density of molecules $n$, \emph{i.e.} $\rho(1)=1/V_0=n$, it follows that:
\begin{equation}
\rho(\Nm) = \frac{\tilde\rho(\Nm)}{\rc^3} =
\frac{\tilde\rho(1)}{\rc^3} = \frac{n}{\Nm},
\end{equation}
since the reduced density is set regardless of CG degree so that $\tilde\rho(\Nm)=\tilde\rho(1)$. Knowing the relation between real and reduced variables, the EOS for the coarse-grained liquid is:
\begin{equation}
p(\Nm) = \frac{n}{\Nm}\kBT + \alpha a \frac{n^2}{\Nm^2}.
\end{equation}
For $\Nm=1$, this simply reduces to the standard form: $p=n \kBT + \alpha a n^2$. 

For a general $\Nm$, we have: $p(\Nm) = \rho(\Nm) \kBT + \alpha a \rho(\Nm)^2$. We now need to decide which quantity is scale invariant. Like F\"uchslin \emph{et al.}, we choose pressure, which is an experimental observable, so $p(\Nm)=p(1)$ for any $\Nm$. As a result, all quantities with the dimension of pressure (\emph{e.g.} compressibility) will be scale invariant. But, in principle, any other variable could be thus chosen. 

To keep pressure scale invariant, the dimension of the ideal gas term dictates that $\kBT$ depends on CG degree as follows:
\begin{equation}
\kBT(\Nm)\sim \Nm.
\end{equation}

% delete
%For a general $\Nm$, we denote all the variables with a prime: $p' = \rho' \kBT' + \alpha' a' \rho'^2$, where $\rho' = \rho(1)/\Nm$.
%Now we need to decide which quantity is scale-invariant. Like F\"uchslin \emph{et al.}, we choose pressure, which is an experimental observable, so $p = p'$. In principle, other variables can be thus chosen. It also follows that
%\begin{equation}
%\kBT'(\Nm)\sim \Nm,
%\end{equation}
%It might seem surprising that to keep the ideal gas term of the EOS scale-invariant, the energy $\kBT$ should depend on the CG degree. To keep the temperature scale-invariant, this implies that the Boltzmann constant $\kB$ must scale linearly with $\Nm$. This is correct, since the dimension of the Boltzmann constant is J/K, and energy was from the very beginning decided to scale linearly, whereas temperature was kept constant.
% end delete

The scaling of the non-ideal term of the EOS has been subject of debates.\cite{Groot_JCP_1997,Fuchslin_JCP_2009} Dimensional analysis reveals that the term $\alpha a$ must scale with $\Nm^2$ to keep pressure scale invariant. To derive the dimension of the fitting constant $\alpha$ and separately the interaction parameter $a$, one can use eq.~(9) from Warren:\cite{Warren_PRE_2003}
\begin{equation}
\alpha = \frac{2\pi}3 \int_0^{\infty} r^3 w(r) dr 
\sim \rc^4(\Nm) \sim \Nm^{4/3}.
\label{eq:alpha}
\end{equation}
Hence, $\alpha$ has the dimension of $\rc^4$ and scales as $\Nm^{4/3}$, and $a$ has the dimension of $\kTc/\rc$ and scales with $\Nm^{2/3}$.

\begin{table}
\begin{ruledtabular}
\begin{tabular}{l|c|c}
Name            & Dimension    & Scaling \\\hline
Length          & $\rc$        & $\Nm^{1/3}$\\
Mass            & $\mc$        & $\Nm$ \\
Number density $\rho$  & $\rc^{-3}$   & $\Nm^{-1}$\\
Energy          & $\kTc$       & $\Nm$  \\
Time            & $(\mc\rc^2/\kTc)^{1/2}$ & $\Nm^{1/3}$ \\
Pressure        & $\kTc/\rc^3$ & 1 \\
Force           & $\kTc/\rc$   & $\Nm^{2/3}$ \\
Parameter $\alpha$         & $\rc^4$    & $\Nm^{4/3}$ \\
Interaction parameter $a$  & $\kTc/\rc$ & $\Nm^{2/3}$
\end{tabular}
\end{ruledtabular}
\caption{Scaling od several quantities expressed in real units with CG degree $\Nm$. In reduced units these quantities are scale invariant.}
\label{tbl:scaling}
\end{table}

Importantly, and as already mentioned, the interaction parameter expressed in reduced units is scale invariant, which can be proved as follows:
%\begin{align}
%\tilde a'& = a' \frac{\rc(\Nm)}{\kTc(\Nm)} \nonumber\\
%&= a \Nm^{2/3} \frac{\rc(\Nm)}{\kTc(\Nm)} =
%\tilde a \frac{\kBT}{\rc} \frac{\rc(\Nm)}{\kTc(\Nm)} \Nm^{2/3} = \tilde a.
%\end{align}
\begin{align}
\tilde a(\Nm) & = a(\Nm) \frac{\rc(\Nm)}{\kTc(\Nm)} \nonumber\\
&= a  \Nm^{2/3} \frac{\rc(\Nm)}{\kTc(\Nm)} \nonumber\\
&= \tilde a \frac{\kTc(1)}{\rc(1)} \frac{\Nm}{\Nm^{1/3}} \frac{\rc(\Nm)}{\kTc(\Nm)} 
= \tilde a.
\end{align}
This is the main and somewhat understated point from the paper by F\"uchslin \emph{et al.}: assuming we do not enforce any constraints from the outside apart from the invariance of the compressibility, \emph{all} the quantities in reduced units remain scale invariant with respect to the coarse-graining. This means that any DPD simulation with water serving as the solvent should be done at $\tilde a=25$. What matters is how we map the results back to the real units after the simulation. This has an important positive consequence in that the interaction parameter does not become too high at high CG degrees, which could lead to freezing, a generally undesirable phenomenon in simulations of liquids.\cite{Trofimov_PHD_2003}

Finally, we derive the scaling of time and the friction constant $\gamma$ from dimensional analysis:
\begin{align}
\tau &= \sqrt{\frac{\mc(\Nm) \rc^2(\Nm)}{\kTc(\Nm)}} \sim 
\sqrt{\frac{\Nm \Nm^{2/3}}{\Nm}} = \Nm^{1/3},\\
\gamma &\sim \frac{\mc(\Nm)}{\tau(\Nm)} \sim \Nm^{2/3}.
\end{align}
F\"uchslin \emph{et al.} wrote that there is a gauge freedom in choosing the scaling of time, but in fact this exponent is determined by the decision to keep pressure scale invariant.

In summary, we have shown that it is possible to use simple dimensional analysis to derive the scaling of quantities in DPD and proved that, in reduced units, all these quantities are scale invariant. The scaling of all the relevant parameters is summarised in Table~\ref{tbl:scaling}.

\subsection{Temperature dependence of interaction parameter}
\label{sec:dpd_temp}
Having explained the scaling with CG degree, we now show that the temperature dependence of the interaction parameter $\tilde a$ due to GW needs to be reconsidered. Noting that the EOS looks essentially the same in real and reduced units: $p=\rho \kBT + \alpha a \rho^2$, we obtain an unambiguous value of the interaction parameter via the matching of compressibility:
\begin{equation}
\kappa^{-1} = \rho\left(\frac{\p p}{\p \rho}\right)_T = \rho \kBT + 2\alpha a \rho^2,
\end{equation}
from which it follows that:
\begin{equation}
a = \frac{\kappa^{-1} - \rho \kBT}{2\alpha\rho^2}.
\label{eq:a}
\end{equation}
So in real units $a$ decreases linearly with temperature.

To convert this equation to reduced units, we simply employ the following relations:
\begin{equation}
p = \tilde p \frac{\kTc}{\rc^3}, \quad
a = \tilde a \frac{\kTc}{\rc}, \quad 
\alpha = \tilde\alpha \,\rc^4, \quad
\rho = \tilde\rho \frac{1}{\rc^3}.
\label{eq:red}
\end{equation}
Note that to non-dimensionalise the compressibility, we cannot use the same approach as GW, who took the molecular density $n=1/V_0$ instead of the DPD density $\rho$. These are only equal to each other in the special case $\Nm=1$. To better illustrate this point and also to expose the strength of the dimensional analysis, we note that inverse compressibility has the same dimension as pressure:
\begin{equation}
\tilde\kappa^{-1} = \kappa^{-1}\frac{\rc^3}{\kTc} = \kappa^{-1} \frac{\tilde\rho V_0}{\kTc}.
\end{equation}
On the other hand, the compressibility due to GW (eq. (14) in their paper) is:\cite{Groot_JCP_1997}
\begin{equation}
\kapnd = \kappa^{-1} \frac{1}{n\kTc} = \kappa^{-1} \frac{V_0}{\kTc}.
\end{equation}\\
These two equations differ by DPD density: $\tilde\kappa^{-1} = \tilde\rho\kapnd$.

Inserting eq.~\eqref{eq:red} into eq.~\eqref{eq:a}, we obtain the interaction parameter in reduced units:
\begin{equation}
\tilde a = \frac{\tilde\kappa^{-1} - \tilde\rho\widetilde{\kBT}}
{2\tilde\alpha \tilde\rho^2},
\end{equation}
and, inserting $\tilde\kappa^{-1} = \tilde\rho\kapnd$ for clarity:
\begin{equation}
\tilde a = \frac{\kapnd - \widetilde{\kBT}}{2\tilde\alpha \tilde\rho}.
\end{equation}
This equation, after setting $\widetilde{\kBT}=1$ and $\kapnd=16$, turns into the form due to GW: $\tilde a = 15/(2\tilde\alpha\tilde\rho)=75/\tilde\rho$. This demonstrates that our derivation based on dimensional analysis is a generalised version of the approach used by GW.

We see that the interaction parameter $\tilde a$ decreases linearly with temperature in reduced units too, as opposed to the linear rise derived by GW, assuming constant compressibility. 
%For a bead containing one water molecule at DPD density $\tilde\rho=3$, the temperature dependence is: $\tilde a = (16 - \widetilde{\kBT})/0.6$.

It must be noted that this temperature dependence is very weak and, for most practical purposes, can be neglected. For example, at 373 K, which is probably the highest temperature at which one would want to simulate liquid water, $\widetilde{\kBT}\approx 1.25$, and $\tilde a$ changes to $\approx 24.6$, which is only a 2\% difference from $\tilde a=25$ at 300 K. However, this variation becomes more relevant if one aims to explore materials at extreme temperatures.

In our analysis so far the compressibility was considered independent of temperature. This might be an overly crude approximation, as, in case of water, the variation is about 10\% between 0 and 50~$^\circ$C.\cite{Fine_JCP_1973} However, the framework presented above enables easy inclusion of this variation by first choosing the energy scale $\kTc$ and simulation temperature $\widetilde{\kBT}$, finding the experimental value of $\kappa$ at the given $\kBT=\widetilde{\kBT}\,\kTc$, and finally non-dimensionalising $\kappa$ with respect to $\kTc$ to obtain $\tilde{\kappa}$.

\section{Rescaling experimental observables}
\label{sec:dpd_rescale_expt}
One of the purposes of DPD is to compute experimental observables of practical interest and compare them with experiment. F\"uchslin \emph{et al.} decided to constrain the three basic units, length, mass, and energy, in such a way that pressure, compressibility, or any other quantity with the same dimension are constant across all the scales. In general, not only liquid compressibility but any experimental observable should be kept constant. However, this is not \emph{a priori} guaranteed by the scaling scheme.

Consider surface tension and self-diffusivity, two important simulation outputs. The dimensional analysis reveals their scale dependence:
\begin{align}
\sigma &\sim \frac{\kTc(\Nm)}{\rc^2(\Nm)} \sim \frac{\Nm}{\Nm^{2/3}} = \Nm^{1/3},\\
D &\sim \frac{\rc(\Nm)^2}{\tau(\Nm)} \sim \frac{\Nm^{2/3}}{\Nm^{1/3}} = \Nm^{1/3}.
\end{align}
Clearly, these experimental observables vary with CG degree, a simulation parameter without physical reality. This is undesirable.

A way to rectify this problem is add an appropriate scaling of the reduced units $\tilde D$ and $\tilde\sigma$ such that these will become scale invariant after conversion to real units. To achieve this, we need to understand how these observables depend on the underlying simulation inputs, such as the interaction parameter $\tilde a$, the Flory-Huggins $\chi$-parameter, or the friction $\tilde\gamma$. To simplify our analysis, we will restrict ourselves to either pure liquids or binary mixtures.

\subsection{Surface tension}
We first turn to the surface tension, which was extensively discussed by Maiti \emph{et al.}\cite{Maiti_JCP_2004} Starting from the Hildebrand solubility parameters $\delta_i$ of species $i$, a simple model for the $\chi$-parameter is:\cite{Barton_book_1990}
\begin{equation}
\chi_{ij} = \frac{V}{\kBT}(\delta_i^2 - \delta_j^2),
\label{eq:chi}
\end{equation}
where $V$ is the bead volume, Maiti \emph{et al.} derived a linear dependence of the $\chi$-parameter on $\Nm$ from the fact that the bead volume varies linearly with the solubilities.\cite{Maiti_JCP_2004}

There are two problems with this line of reasoning: a technical one and a theoretical one. Technically, these authors kept the energy scale $\widetilde{\kBT}$ invariant. If we corrected this, we would find out that $\chi$ is invariant, which would lead, together with an invariant $\tilde a$, to an invariant surface tension $\tilde\sigma$. However, this would imply the scale dependence of $\sigma$.

The theoretical objection is that mixing is a delicate interplay of various effects on the microscale and it is not \emph{a priori} clear how these should vary with the number of molecules incorporated into a bead. The coarse-graining is in itself an artificial process without any physical reality, the sole aim of which is speeding up the simulation.

In order to derive a plausible scaling of $\tilde\sigma$, we follow a different route, which will not require diving into the complex microscopic origin of mixing. We start from the dependence of surface tension on the $\chi$-parameter derived by GW in the context of the DPD (eq. (36) in their paper, with $\tilde\rho$ being density and assuming $\kTc=1,\rc=1$):\cite{Groot_JCP_1997}
\begin{equation}
\tilde\sigma = \begin{cases}
0.75 \tilde\rho \chi^{0.26} \bigg(1 - \frac{2.36}{\chi}\bigg)^{3/2} & \chi > 2.36 \\
0, & \chi \leq 2.36.
\end{cases}
\label{eq:sigma_chi}
\end{equation}
To render $\tilde\sigma$ scale invariant, we determine the scaling of the $\chi$-parameter such that $\tilde\sigma \sim \Nm^{-1/3}$. In other words, we are looking for the exponent $\beta$ such that:
\begin{align}
\sigma &= \tilde\sigma\frac{\kTc(\Nm)}{\rc^2(\Nm)}  \\
&= 0.75\tilde\rho(\chi\Nm^\beta)^{0.26}
\bigg(1-\frac{2.36}{\chi\Nm^\beta}\bigg)^{3/2} \frac{\kTc(1)}{\rc^2(1)}
\frac{\Nm}{\Nm^{2/3}}\\
&\sim \text{constant}.\nonumber
\end{align}
Due to the rather complex power law of eq.~\eqref{eq:sigma_chi}, we resort to numerical minimisation after defining the relevant range of CG degrees. Although it might be desirable to try to deliver a perfect analytical solution, given the overall qualitative nature of the DPD, a reasonably accurate approximation is sufficient for practical simulations.

We consider the mixtures explored by Maiti, that is water--benzene, water--CCl$_4$, and water--octane. Their $\chi$-parameters are computed from the Hildebrand solubilities, and the data are summarised in Table~\ref{tbl:liquids}. Defining the range of CG degrees $\Nm\in\{1,2,...,10\}$ and the root mean-square error:
\begin{equation}
\text{RMSE}=\sqrt{\frac{1}{N_{\Nm}-1} \sum_{\Nm} (\sigma_1-\sigma_{\Nm})^2},
\label{eq:rmse}
\end{equation}
we can minimise the RMSE across these mixtures. Hence, we arrive at the scaling of the $\chi$-parameter $\chi\sim\Nm^{-0.22}$.

\begin{table}
\centering
\begin{ruledtabular}
\begin{tabular}{lccc}\toprule
Component    & $\delta$(MPa$^{1/2}$) & $\chi$ & $\sigma_{\rm{expt}}$ (mN/m)\\\hline
Water        & 47.9 &       & \\
\rm{Benzene} & 18.6 & 6.132 & 35.0 \\
\rm{CCl$_4$} & 17.8 & 6.474 & 45.0 \\
\rm{Octane}  & 15.6 & 7.555 & 51.7 \\\bottomrule
\end{tabular}
\end{ruledtabular}
\caption{Solubilities $\delta$, $\chi$-parameters and surface tensions $\sigma_{\rm{expt}}$ of water--liquid interface, taken from Maiti \emph{et al.}\cite{Maiti_JCP_2004}}
\label{tbl:liquids}
\end{table}

\begin{figure*}
\centering
\begin{subfigure}[b]{0.45\textwidth}
\includegraphics[width=1\textwidth]{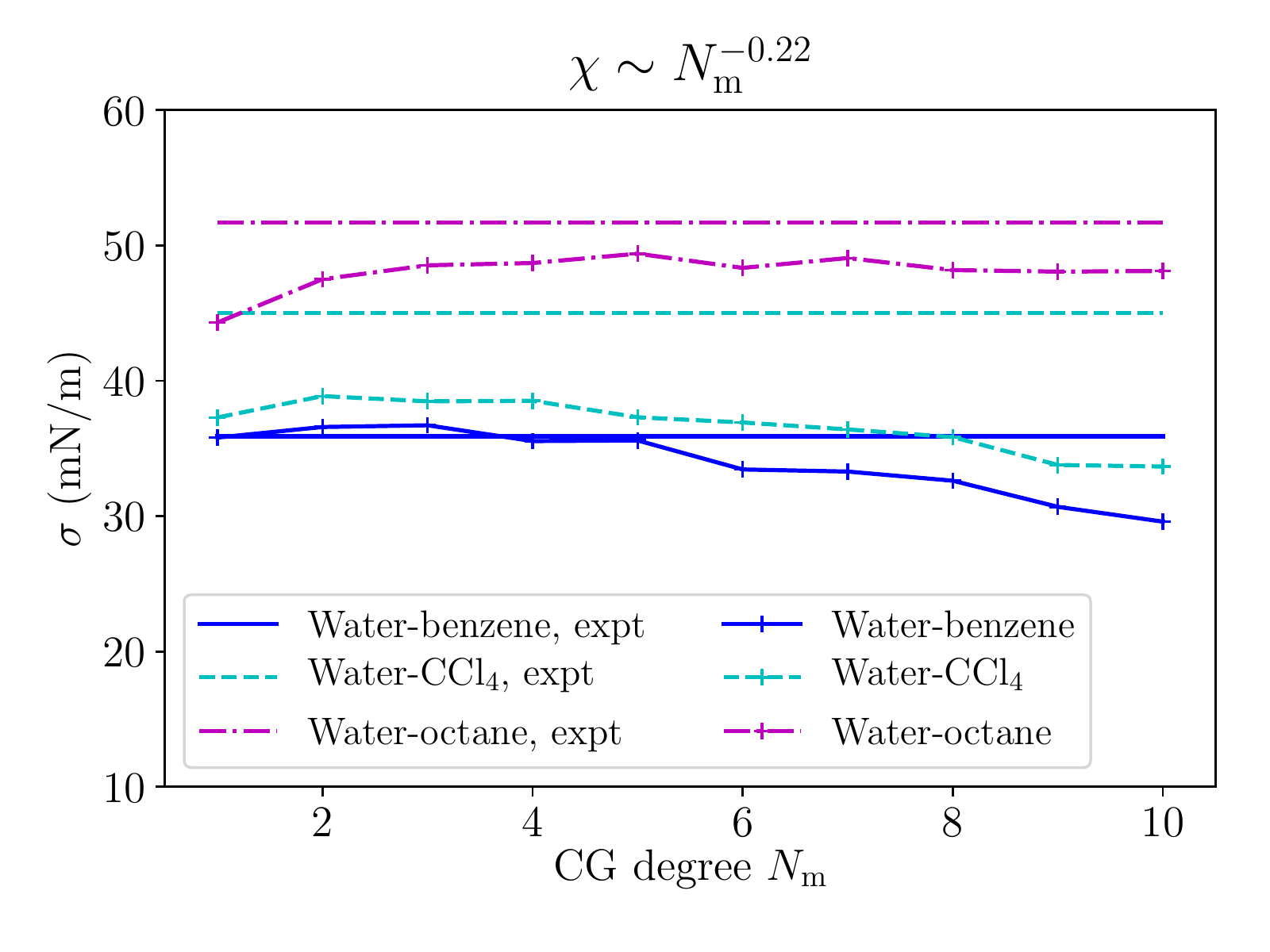}
\caption{}
\end{subfigure}
~
\begin{subfigure}[b]{0.45\textwidth}
\includegraphics[width=1\textwidth]{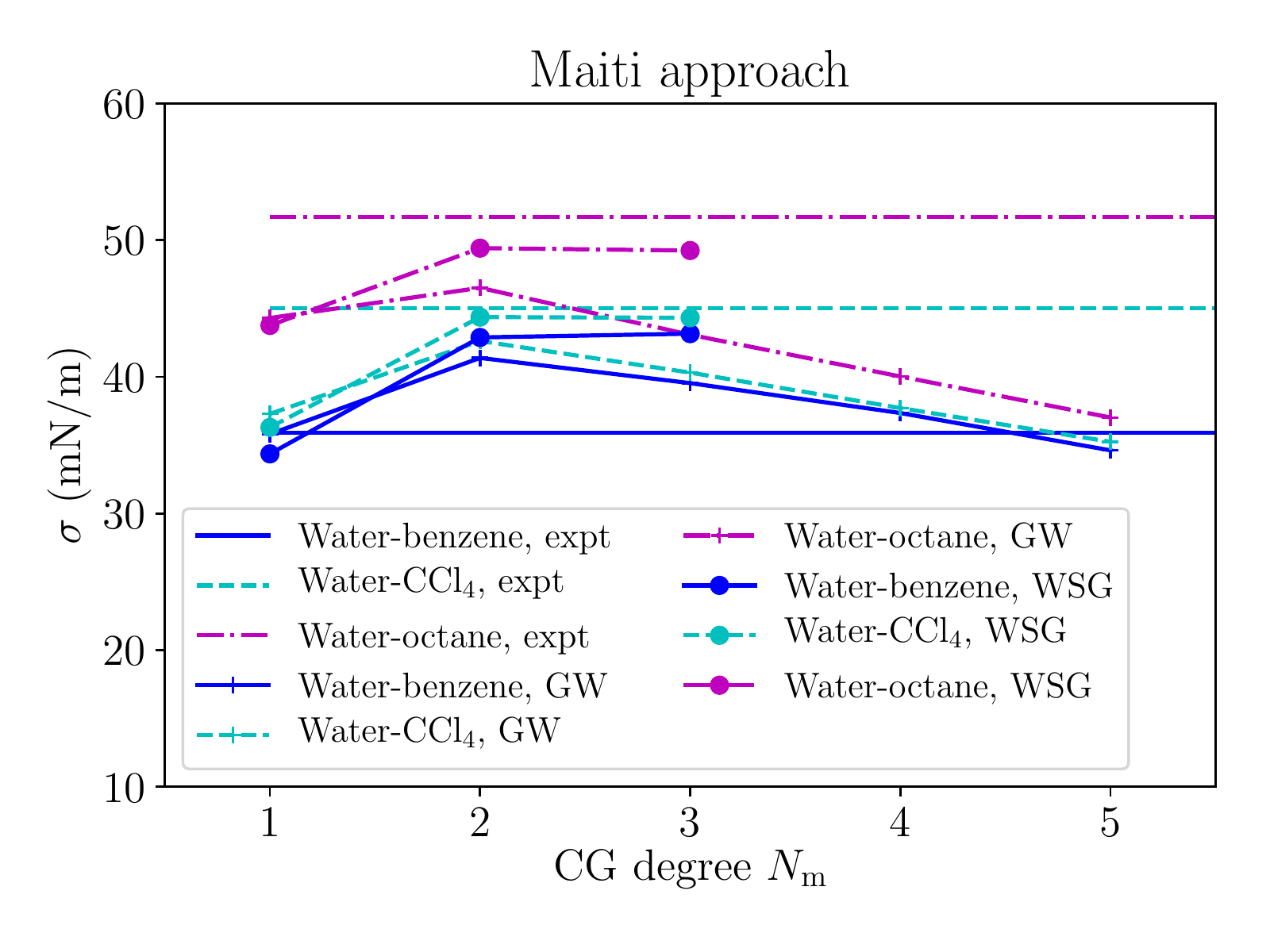}
\caption{}
\end{subfigure}
\caption{Variation of the surface tension for three mixtures with coarse-graining degree: (a) our derivation using the $\Nm^{-0.22}$ scaling of $\chi$-parameter, and (b) methodology by Maiti,\cite{Maiti_JCP_2004} working up to $\Nm=5$, for both Groot \& Warren (GW) and Wijmans, Smit and Groot (WSG) $\Delta a-\chi$ relation.}
\label{fig:sigmas}
\end{figure*}

To test this scaling, we performed simulations with the LAMMPS software package.\cite{LAMMPS} We set a $20\times 10\times 10$ orthorhombic cell at density $\tilde\rho=3$. The time step was set to 0.05. Taking water as the default liquid, the volume of a bead containing one molecule was $V_0=30$~\AA, and the bead self-repulsion was $\tilde a_{ii}=25$. We equilibrated the system for 20k steps and collected data for another 50k steps. The surface tension was calculated from the pressure tensor components:
\begin{equation}
\tilde\sigma = \frac{\tilde L_x}{2} \bigg(\langle\tilde p_{xx}\rangle - 
\frac{\langle\tilde p_{yy}\rangle + \langle\tilde p_{zz}\rangle} 2 \bigg).
\label{eq:sigma}
\end{equation}

In parallel, we have reproduced the measurements by Maiti \emph{et al.} These workers tested two various relations for $\Delta a$ vs $\chi$: a linear one derived by GW:
\begin{equation}
\Delta a = \chi/0.286,
\end{equation}
which we used for our simulations as well, and a quadratic one derived by Wijmans \emph{et al.} (WSG):\cite{Wijmans_JCP_2001}
\begin{equation}
\frac{\chi}{\Delta a} = 0.3 - \frac{0.3-0.2}{115-15}(\Delta a - 15).
\end{equation}
Both of these, if scaled linearly with $\Nm$, lead to extremely large excess repulsions $\Delta a$ and allow CG degrees only up to $\Nm=5$ and 3, respectively.

Fig.~\ref{fig:sigmas} shows the results of analytical predictions and simulations using the scaling arguments presented above, and the approach by Maiti \emph{et al.} Our method gives satisfactory results for water--benzene and water--octane mixtures for CG degrees up to $\Nm=10$ and possibly even above. The water--CCl$_4$ mixture starts from an incorrect position at $\Nm=1$, which might be due to the inability of the overly simple eq.~\eqref{eq:chi} to describe real behaviour. Overall, our derived scaling of $\chi\sim\Nm^{-0.22}$ is able to capture the mixing properties over a wide range of CG degrees and improve the predictive accuracy of mixing in DPD.

\begin{figure*}
\centering
\begin{subfigure}[b]{0.45\textwidth}
\includegraphics[width=1\textwidth]{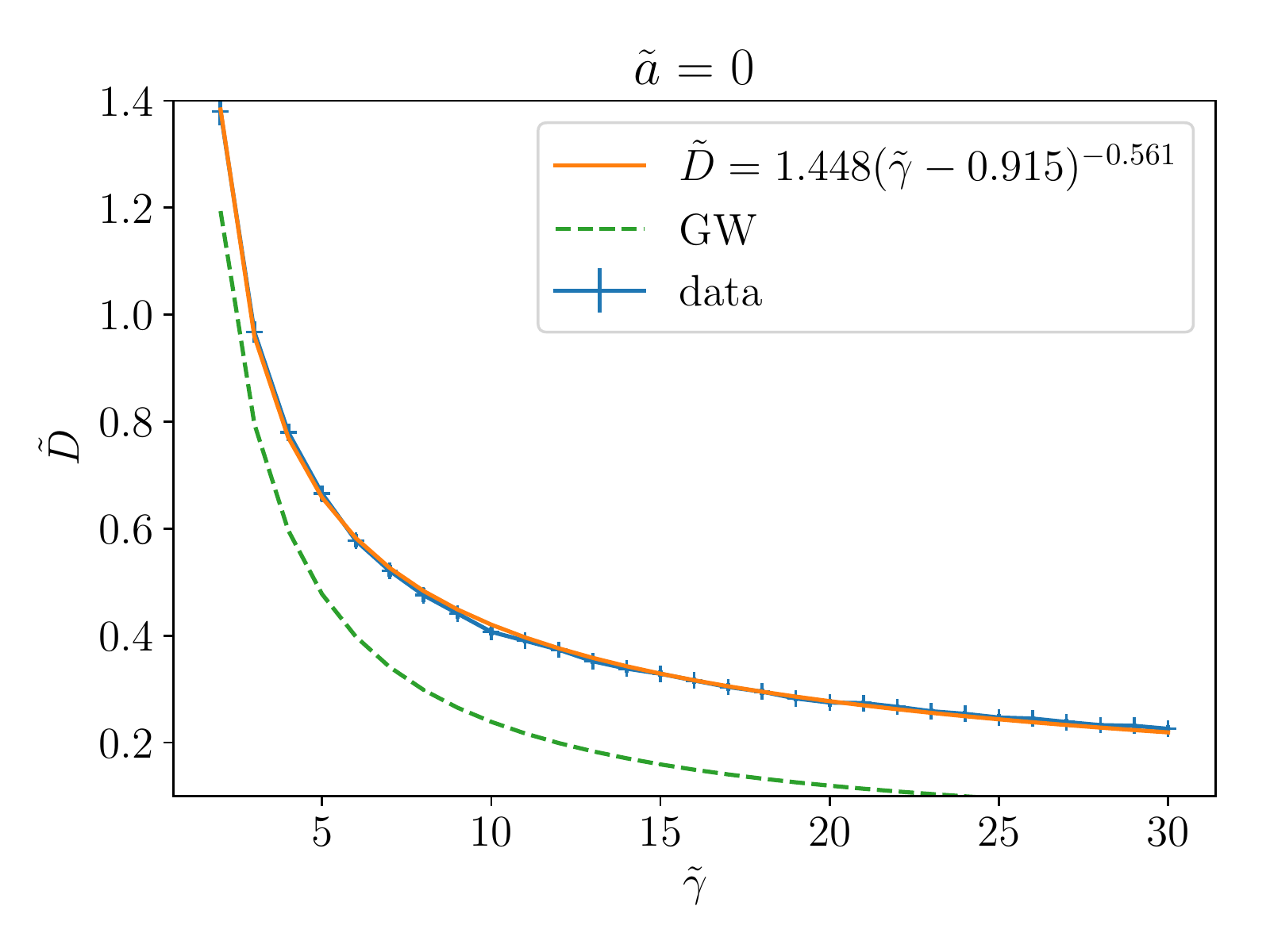}
\caption{}
\end{subfigure}
~
\begin{subfigure}[b]{0.45\textwidth}
\includegraphics[width=1\textwidth]{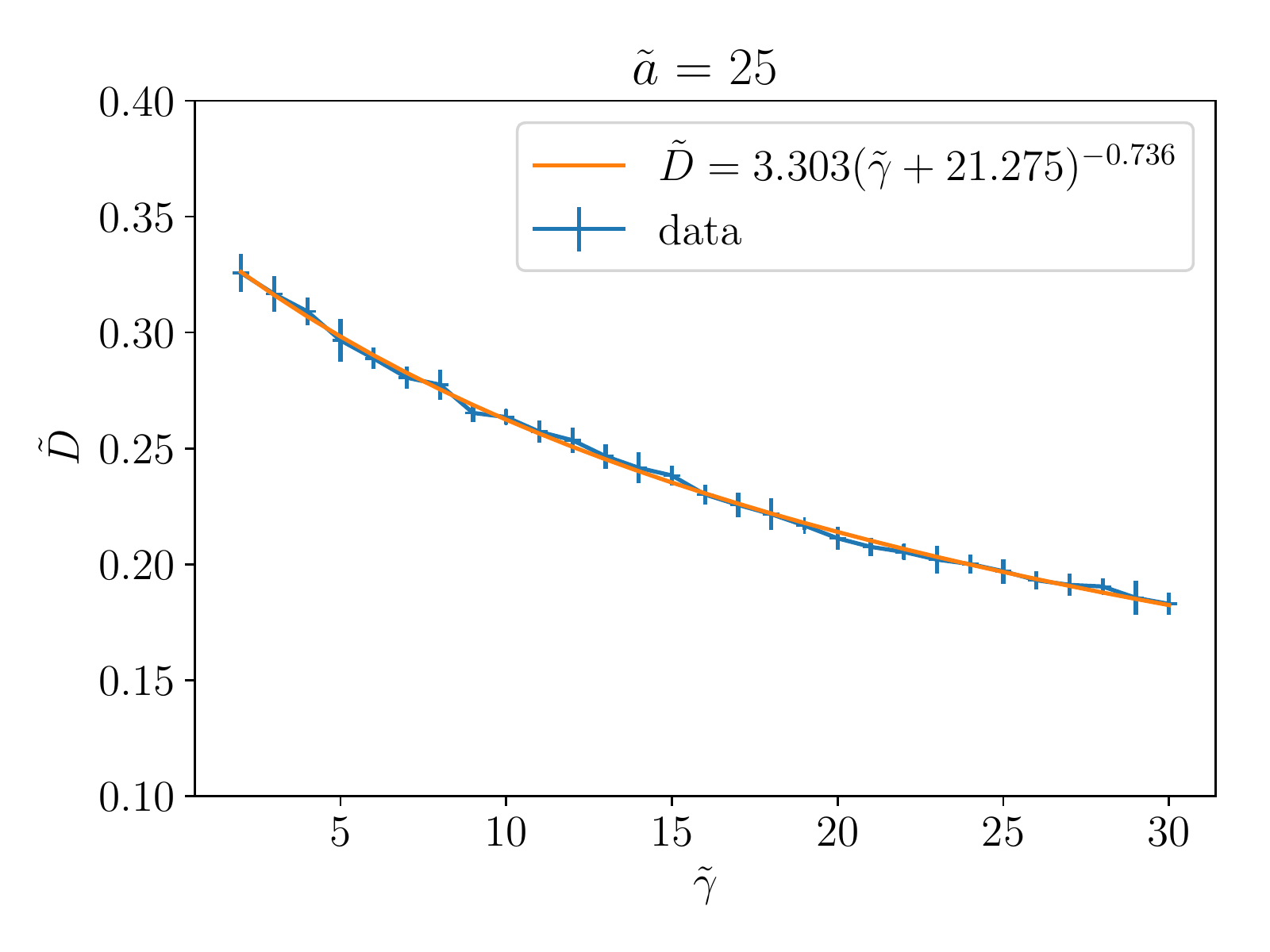}
\caption{}
\end{subfigure}
\caption{Self-diffusivity dependence with error bars for (a) $\tilde a=0$, \emph{i.e.} no conservative interaction, and (b) $\tilde a=25$ with power law fits (equations shown in inset).}
\label{fig:d_a}
\end{figure*}

\subsection{Self-diffusivity}
The friction parameter $\gamma$ from the dissipative and random force (eq.~\eqref{eq:fd}) is a microscale parameter that influences the self-diffusivity $D$, which can be measured experimentally. Overall, bead diffusion depends not only on $\gamma$, but also on the particle repulsion $a$. There have been attempts to analytically derive how $\gamma$ should vary with CG degree.\cite{Izvekov_JCP_2014} Nonetheless, we can easily derive the scaling that renders the self-diffusivity constant across different CG degrees. 

To understand the behaviour of a pure DPD liquid, we exploit the fact that the low number of simulation parameters enables fast exploration of a large portion of the parameter space. Defining the self-diffusivity from the MSD:
\begin{equation}
\tilde D = \lim_{\tilde t\rightarrow\infty} \frac{\tilde r(\tilde t)^2}{6\tilde t},
\quad D = \tilde D \frac{\rc^2}{\tau},
\end{equation}
we measured the dependence of $\tilde D$ for a wide range of $\tilde\gamma$ values between 2 and 30, and $\tilde a$ values between 0 and 55. Using a $10\times 10\times 10$ orthorhombic cell with 3000 beads, we equilibrated for 40k time steps and measured the MSDs for 1000 steps 10 times in succession to eliminate noise. We took a smaller time step 0.03 to maintain the temperature at $\kTc=1$, as it tends to diverge with increased friction.

Firstly, we consider the case where $\tilde a=0$, \emph{i.e.} beads interact only via a dissipative and random force. In this case, using a mean-field approximation by setting $g(r)=1$, GW derived analytically $D=45/(2\pi\gamma\rho\rc^3)$, or, in reduced units, $\tilde D=45/(2\pi\tilde\gamma\tilde\rho)$.\cite{Groot_JCP_1997} However, from simulations we obtained systematically higher values, as shown on Fig.~\ref{fig:d_a} (left). For all the interaction parameters $\tilde a$, it is possible to fit the self-diffusivity with the power law of the form:
\begin{equation}
\tilde D(\tilde\gamma) = c_1(\tilde\gamma - c_2)^{c_3},
\end{equation}
where $c_i,i\in\{1,2,3\}$ are fitting parameters. We also tried to fit the self-diffusivities for both $a$ and $\gamma$ at once via:
\begin{equation}
\tilde D(\tilde\gamma,\tilde a) = c_1(\tilde\gamma - c_2 \tilde a)^{c_3},
\end{equation}
but this failed to achieve good accuracy, especially at low frictions. This is not an important obstacle, since most simulations are done with water as the default bead type with the repulsion $\tilde a=25$. Hence, to derive the scaling of $\tilde\gamma$ with the CG degree, it is sufficient to focus only on this value.

\begin{figure}
\centering
\includegraphics[width=0.45\textwidth]{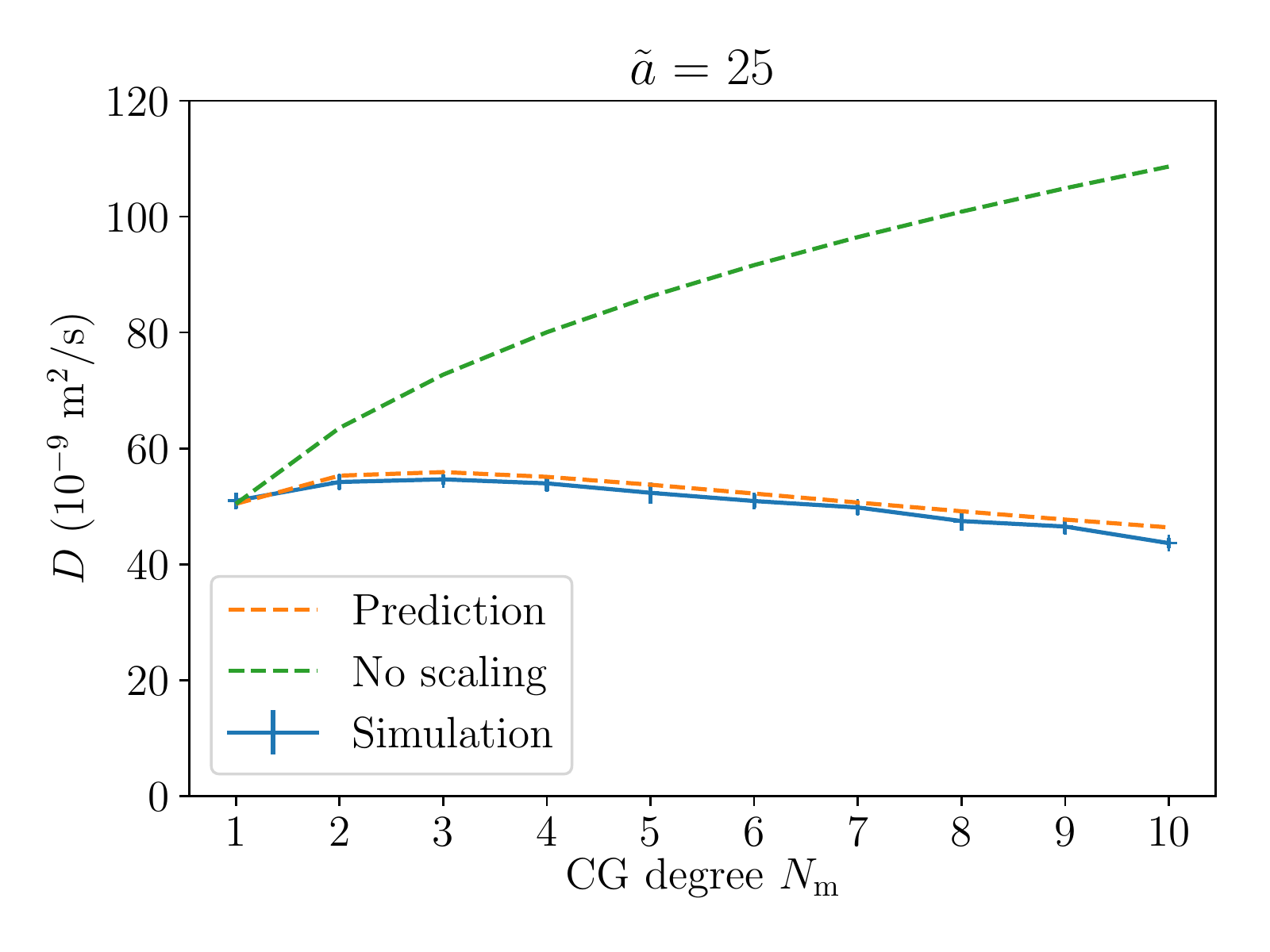}
\caption{Self-diffusivity of DPD water (in real units) as a function of the CG degree, performed at $\tilde a=25$ and $\tilde\gamma=4.5$ and plotted with or without the offsetting scaling of $\tilde\gamma$ with CG degree. Simulation results show agreement with the scaling $\tilde\gamma\sim\Nm^{1.13}$ predicted from eq.~\eqref{eq:diff_sc}.}
\label{fig:verify}
\end{figure}

As before with the surface tension, our aim is to obtain the exponent $\beta$ such that:
\begin{align}
D &= \tilde D \frac{\rc^2(\Nm)}{\tau(\Nm)} \\
&= 3.303(\tilde\gamma\Nm^\beta + 21.275)^{-0.736} \frac{\rc^2(1)}{\tau(1)}\frac{\Nm^{2/3}}{\Nm^{1/3}} \sim \text{const}.%\nonumber
\label{eq:diff_sc}
\end{align}
Starting from $\tilde\gamma(1)=4.5$ at $\Nm=1$ used by GW, we have minimised the RMSE defined as in eq.~\eqref{eq:rmse} for $\Nm\in\{1,...,10\}$, and obtained $\beta=1.13$. To verify this, we have again simulated pure liquids at $\tilde a=25$ with $\tilde\gamma(\Nm)=\tilde\gamma(1)\Nm^{1.13}$. The results on Fig.~\ref{fig:verify} show a reasonably, if not perfectly flat curve, demonstrating the achieved scale invariance of water self-diffusivity in DPD.

Compared with the experimental self-diffusivity of water $2.3\times 10^{-9}$~m$^2$/s at 300~K, the values from DPD simulations are about 20 times larger. This is expected due to the extremely soft nature of the DPD potential. To precisely target the experiment, we would need to take $\tilde\gamma$ of about 1500. Such a large value would severely impact the simulation efficiency in that the time step would have to be orders of magnitude smaller, and the speed of equilibration, which is one of the principal advantages of the DPD, would be lost. Nonetheless, having a method to generate scale invariant, if shifted self-diffusivities can improve the insight into the dynamics of soft matter.

% =====
% many-body dpd
% =====
\section{Mixing in many-body DPD}
\label{sec:mdpd}
\subsection{Overview of MDPD}
Having understood the scaling of standard DPD, we now turn to its many-body version. First presented by Pagonabarraga \emph{et al.} and Trofimov \emph{et al.},\cite{Pagonabarraga_JCP_2001, Trofimov_JCP_2002} and thoroughly explored by Warren,\cite{Warren_PRE_2003} MDPD builds on top of standard DPD by adding a density-dependent interaction at a new length scale $\trd<1$. This modification leads to an EOS with a van der Waals loop, which enables the formation of a liquid-vapour interface and increases the applicability to free surfaces. Compared with standard DPD, which contains only repulsive interactions, MDPD can support simulations in which the bead density varies widely across the simulation cell.

Adopting reduced units as in Section~\ref{sec:dpd}, the full MDPD force field is:
\begin{equation}
\vc{\tilde F}_{ij}(\vc{\tilde r}) = \tilde A w(\tilde r) \vc{\hat r} + 
\tilde B(\bar\rho_i + \bar\rho_j) \wwd (\tilde r) \vc{\hat r},
\label{eq:ff}
\end{equation}
where $\tilde A$ and $\tilde B$ are interaction parameters, $\tilde r=|\vc{\tilde r}|$, $\vc{\hat r} = \vc{\tilde r}/\tilde r$. $\wwd(\tilde r)$ is a modified weight function:
\begin{equation}
\wwd(\tilde r) = \begin{cases}
1 - \tilde r/\trd, & \tilde r \leq \trd \\
0,                 & \tilde r > \trd.
\end{cases}
\end{equation}
The local density $\bar\rho_i$ of particle $i$ is defined as:
\begin{equation}
\bar\rho_i = \sum_{j\neq i} w_\rho(\tilde r_{ij}),
\end{equation}
where:
\begin{equation}
w_{\rho}(\tilde r) = \begin{cases}
\frac{15}{2\pi\trd^3} \left(1 - \frac{\tilde r}{\trd}\right)^2, 
& \tilde r \leq \trd,\\
0, & \tilde r >\trd.
\end{cases}
\end{equation}
We stress that the index $j$ runs over \emph{all} the particles, not just those of the same type as $i$th particle.

Warren showed that, for $\tilde A<0$ and $\tilde B>0$, this force field leads to the liquid-vapour coexistence, and determined its equation of state:\cite{Warren_PRE_2003}
\begin{equation}
\tilde p = \tilde\rho
+ \tilde\alpha \tilde A \tilde\rho^2 
+ 2\talphaMB \tilde B
(\tilde\rho^3 - \tilde c\tilde\rho^2 + \tilde d),
\label{eq:eos_w}
\end{equation}
where $\tilde\alpha=0.1$ comes from standard DPD, $\alphaMB = \int_0^{\infty} r^3 \wwd(r) dr = \pi\trd^4/30\approx 0.1\trd^4$, and $\tilde c=4.16$, $\tilde d=18$ are fitting constants. This EOS was further improved by Jamali \emph{et al.}:\cite{Jamali_JCP_2015}
\begin{equation}
\tilde p = \tilde\rho + 
\tilde\alpha \tilde A \tilde\rho^2 
+ 2\talphaMB \tilde B
(\tilde\rho^3 - \tilde c' \tilde\rho^2 + \tilde d' \tilde\rho) 
-\frac{\talphaMB \tilde B}{|\tilde A|^{1/2}}\tilde\rho^2,
\label{eq:eos_j}
\end{equation}
where $\tilde c'=4.69$ and $\tilde d'=7.55$. For further work, we decided to use the more accurate version of the EOS due to Jamali \emph{et al.}

In the regime of the liquid-vapour coexistence, we can derive how the density and surface tension depend on the parameters $\tilde A, \tilde B,\trd$, and, by inverting thus obtained relations, determine $\tilde A$ and $\tilde B$ that would enable the simulation of a real liquid with a given experimental surface tension.

\subsection{Parameterisation for real liquids}
In our recent work\cite{Vanya_PRE_2018} we determined the regions of the phase diagram of an MDPD fluid that give rise to the liquid phase. Based on the measurements of liquid density and surface tension as a function of the interaction parameters $\tilde A,\tilde B$ and fixing $\trd$, for example at 0.75, we solved for the interaction parameters from the material properties, in this case compressibility and surface tension.

For any liquid defined by compressibility, surface tension and volume per molecule, and choosing CG degree $\Nm$ and temperature defining the energy scale $\kTc$, we have four highly non-linear equations with four unknowns: $\rc, \tilde A,\tilde B$, and $\tilde\rho$. Considering, \emph{e.g.}, $\trd=0.75$, the fitting coefficients from Table I and II in Ref.\cite{Vanya_PRE_2018} yield:

\begin{widetext}
\begin{align}
\rc &= (\tilde\rho\Nm V_0)^{1/3}, \\
\tilde\rho(\tilde A,\tilde B)   &= 3.01 + 1.21 (-\tilde A) \tilde B^{-0.856},\\
\tilde\sigma(\tilde A,\tilde B) &= \sigma \frac{\rc^2}{\kTc}
= (0.0807 \tilde A^2 + 0.526 \tilde A) (\tilde B + 0.0659\tilde A)^{-0.849},\\
\tilde\kappa^{-1} &= \kappa^{-1} \frac{\kTc}{\rc^3} =
  \tilde\rho \frac{\partial{\tilde p}}{\partial{\tilde\rho}} =
  \tilde\rho + 2\tilde\alpha\tilde A\tilde\rho^2 + 
 2\talphaMB\tilde B (3\tilde\rho^3 - 2c'\tilde\rho^2 + d'\rho) 
- \frac{\talphaMB\tilde B}{|\tilde A|^{1/2}} 2\tilde\rho^2.
\label{eq:4eqs}
\end{align}
\end{widetext}

These equations can be solved numerically, either by a root-finding algorithm or by a brute-force search through the parameter space.

In a mesoscale simulation, one does not demand extreme accuracy and rounding the interaction parameters to only a few decimal places is sufficient. Hence, working with resolution $\delta\tilde A=0.1$, $\delta\tilde B=0.1$, a brute-force search through the parameter space with range $[-100, 0]$ and $[0,100]$ for $\tilde A$ and $\tilde B=1$, respectively, requires only about 10k evaluations of eqs.~\eqref{eq:4eqs} and an objective error term. On an average modern computer, this process takes a few seconds.

We defined the error function as follows:
\begin{equation}
\text{Err} = w\left| 1-\frac{\sigma}{\sigma_{\rm expt}}\right| + 
\left| 1 - \frac{\kappa^{-1}}{\kappa^{-1}_{\rm expt}}\right|,
\end{equation}
where $\sigma_{\rm expt}$ and $\kappa^{-1}_{\rm expt}$ are experimental surface tension and compressibility, respectively. We chose the weight factor $w=5$, putting more emphasis on reproducing surface tension more accurately than compressibility, since the latter is in itself too restrictive, as has been recently highlighted in the context of standard DPD.\cite{Anderson_JCP_2017,Fraaije_JCIM_2016}

We have determined the interaction parameters $\tilde A, \tilde B$ for water, which we later apply to water-solvent mixtures. We need to bear in mind that water is an outlier in that its surface tension is about three times higher and the volume per molecule is several times lower than in case of other common solvents. We have explored a range of many-body cutoffs $\trd$: 0.65, 0.75 and 0.85 and CG degrees $\Nm$ from 1 to 10.

The resulting values of $\tilde A,\tilde B$ for $\trd=0.65$, which are shown in Table~\ref{tbl:sol_65}, are relatively small and marked by excessive inverse compressibilities. More importantly, the reduced density, which is a key parameter for simulation efficiency, is extremely high for any CG degree up to 10, as can be compared by the typical density $\tilde\rho=3$ used in standard DPD. We conclude that this many-body cutoff is useless for water simulations and decide not to proceed.

The parameter search for $\trd=0.75$ yields more suitable results, with accurate surface tensions as well as compressibilities for all CG degrees, as shown in Table~\ref{tbl:sol_75}. The density $\tilde\rho$ is still rather high at $\Nm=1$ and 2, but other CG degrees are viable. $\trd=0.85$ in Table~\ref{tbl:sol_85} produces reasonable parameter values and highly suitable reduced densities, almost at the level of standard DPD, but slightly low inverse compressibilities. Hence, both of these values of $\trd$ are suitable for simulations including water. This analysis also suggests that an intermediate value of $\trd$, such as 0.80, would provide both reasonable densities as well as accurate compressibilities.

\begin{table}
\centering
\begin{ruledtabular}
\begin{tabular}{cccccc}\toprule
$\Nm$ & $\tilde\rho$ & $\tilde A$ & $\tilde B$ & $\sigma$ (mN/m) & 
$\kappa^{-1}$ (10$^9$~Pa)\\\hline
1 & 22.10 & $-14.8$ & 2.0 & 71.1 & 3.52 \\
2 & 21.52 & $-14.3$ & 2.0 & 71.1 & 3.37 \\
3 & 20.61 & $-14.1$ & 2.1 & 71.8 & 3.33 \\
4 & 21.06 & $-13.9$ & 2.0 & 71.8 & 3.24 \\
5 & 20.28 & $-13.8$ & 2.1 & 71.3 & 3.23 \\
6 & 20.83 & $-13.7$ & 2.0 & 72.2 & 3.18 \\
7 & 20.71 & $-13.6$ & 2.0 & 70.8 & 3.15 \\
8 & 19.46 & $-13.6$ & 2.2 & 71.5 & 3.20 \\
9 & 20.60 & $-13.5$ & 2.0 & 71.4 & 3.12 \\
10 & 19.36 & $-13.5$ & 2.2 & 71.4 & 3.17 \\
\bottomrule
\end{tabular}
\end{ruledtabular}
\caption{Derived interaction parameters for water at various CG degrees and $\trd=0.65$.}
\label{tbl:sol_65}
\end{table}

\begin{table}
\centering
\begin{ruledtabular}
\begin{tabular}{cccccc}\toprule
$\Nm$ & $\tilde\rho$ & $\tilde A$ & $\tilde B$ & $\sigma$ (mN/m) & 
$\kappa^{-1}$ (10$^9$~Pa)\\\hline
1 & 9.99 & $-18.5$ & 3.9 & 71.6 & 2.23 \\
2 & 8.63 & $-18.1$ & 4.9 & 71.5 & 2.16 \\
3 & 7.76 & $-18.2$ & 6.0 & 71.5 & 2.19 \\
4 & 7.23 & $-18.2$ & 6.9 & 71.3 & 2.22 \\
5 & 6.94 & $-18.0$ & 7.4 & 71.4 & 2.20 \\
6 & 6.70 & $-17.9$ & 7.9 & 71.6 & 2.20 \\
7 & 6.55 & $-17.7$ & 8.2 & 71.5 & 2.18 \\
8 & 6.39 & $-17.6$ & 8.6 & 71.4 & 2.18 \\
9 & 6.23 & $-17.6$ & 9.1 & 71.5 & 2.20 \\
10 & 6.12 & $-17.5$ & 9.4 & 71.5 & 2.20 \\
\bottomrule
\end{tabular}
\end{ruledtabular}
\caption{Derived interaction parameters for water at various CG degrees and $\trd=0.75$.}
\label{tbl:sol_75}
\end{table}

\begin{table}
\centering
\begin{ruledtabular}
\begin{tabular}{cccccc}\toprule
$\Nm$ & $\tilde\rho$ & $\tilde A$ & $\tilde B$ & $\sigma$ (mN/m) & 
$\kappa^{-1}$ (10$^9$~Pa)\\\hline
1 & 5.71 & $-39.8$ & 10.0 & 71.3 & 1.20 \\
2 & 5.43 & $-39.9$ & 11.0 & 71.6 & 1.16 \\
3 & 5.24 & $-39.6$ & 11.6 & 71.5 & 1.10 \\
4 & 5.07 & $-40.0$ & 12.5 & 71.4 & 1.09 \\
5 & 4.95 & $-40.0$ & 13.1 & 71.4 & 1.06 \\
6 & 4.88 & $-39.6$ & 13.3 & 71.5 & 1.01 \\
7 & 4.80 & $-39.4$ & 13.6 & 71.4 & 0.98 \\
8 & 4.68 & $-40.0$ & 14.6 & 71.6 & 0.99 \\
9 & 4.60 & $-40.0$ & 15.1 & 71.3 & 0.96 \\
10 & 4.54 & $-40.0$ & 15.5 & 71.5 & 0.94 \\
\bottomrule
\end{tabular}
\end{ruledtabular}
\caption{Derived interaction parameters for water at various CG degrees and $\trd=0.85$.}
\label{tbl:sol_85}
\end{table}

\begin{figure}
\centering
\includegraphics[width=0.45\textwidth]{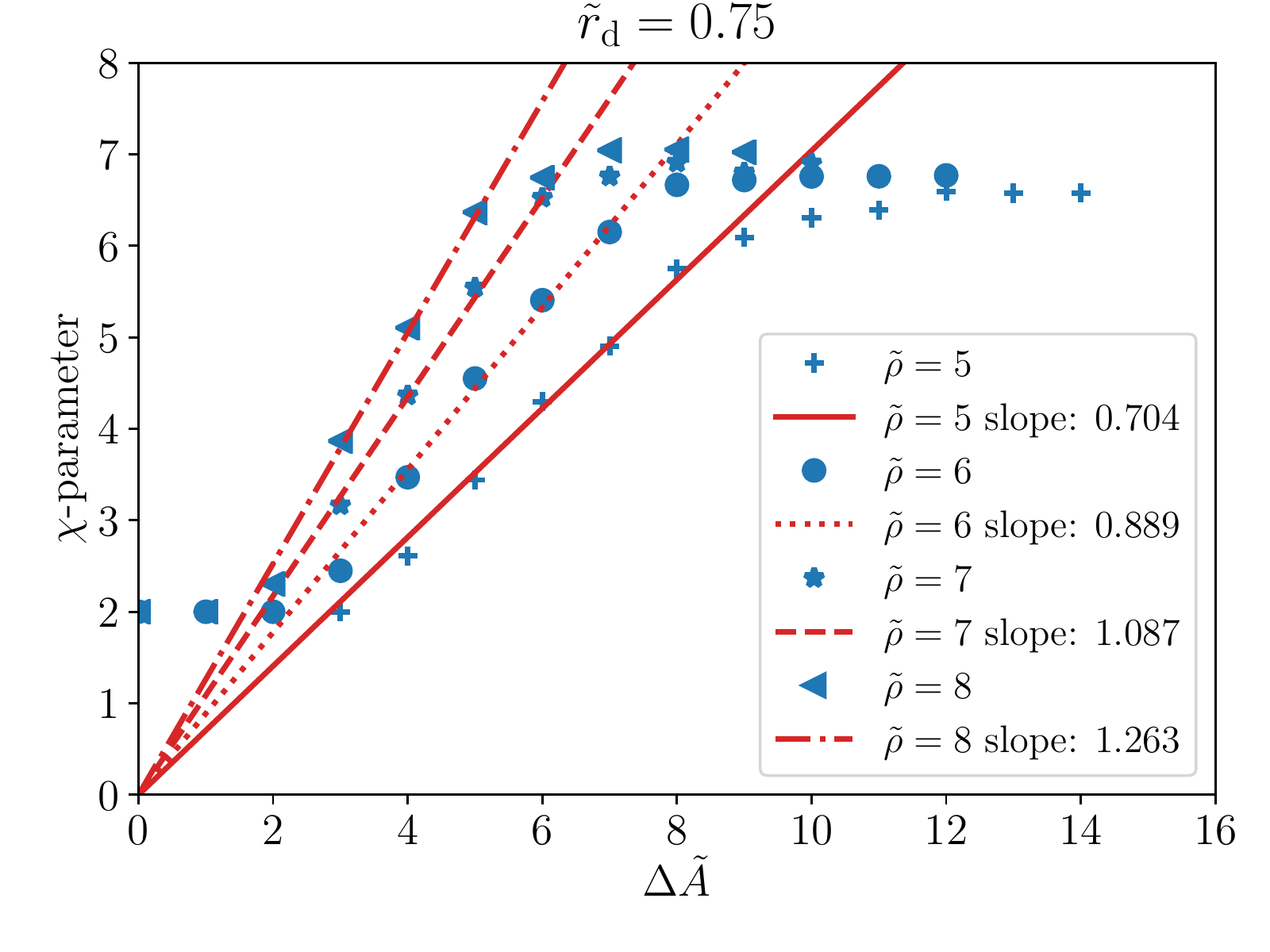}
\caption{Dependence of the Flory-Huggins $\chi$-parameter on excess repulsion $\Delta\tilde A$ for a range of densities. Strong deviation from the linear regime at low and high values of $\Delta\tilde A$ is revealed here, in comparison with Fig.~6 from GW\cite{Groot_JCP_1997} or Fig.~10 from Jamali \emph{et al.}\cite{Jamali_JCP_2015}}
\label{fig:chi_da}
\end{figure}

\begin{figure*}[t!]
\centering
\begin{subfigure}[b]{0.45\textwidth}
\includegraphics[width=1\textwidth]{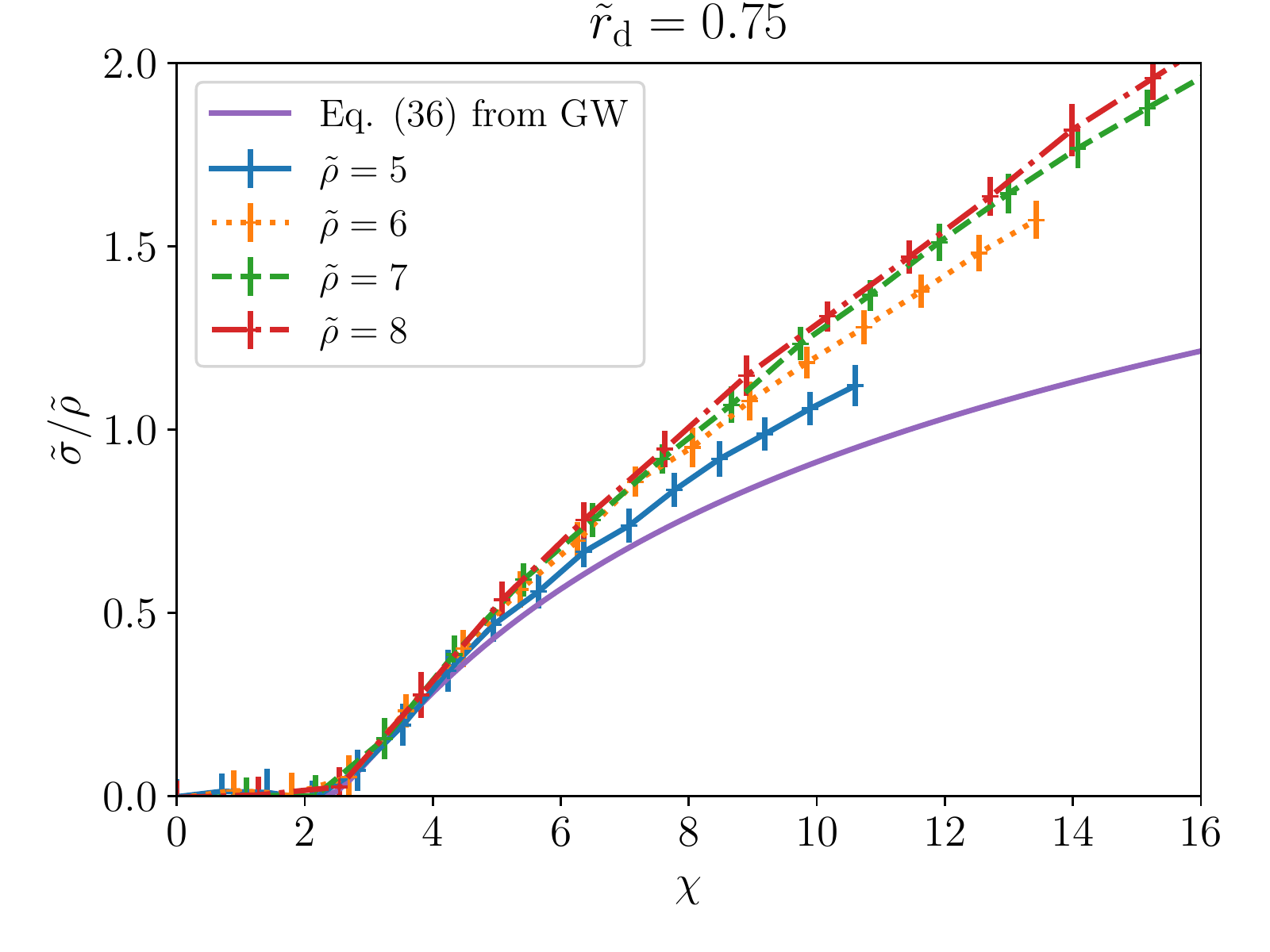}
\caption{}
\label{fig:sigma_rd_75}
\end{subfigure}
~
\begin{subfigure}[b]{0.45\textwidth}
\includegraphics[width=1\textwidth]{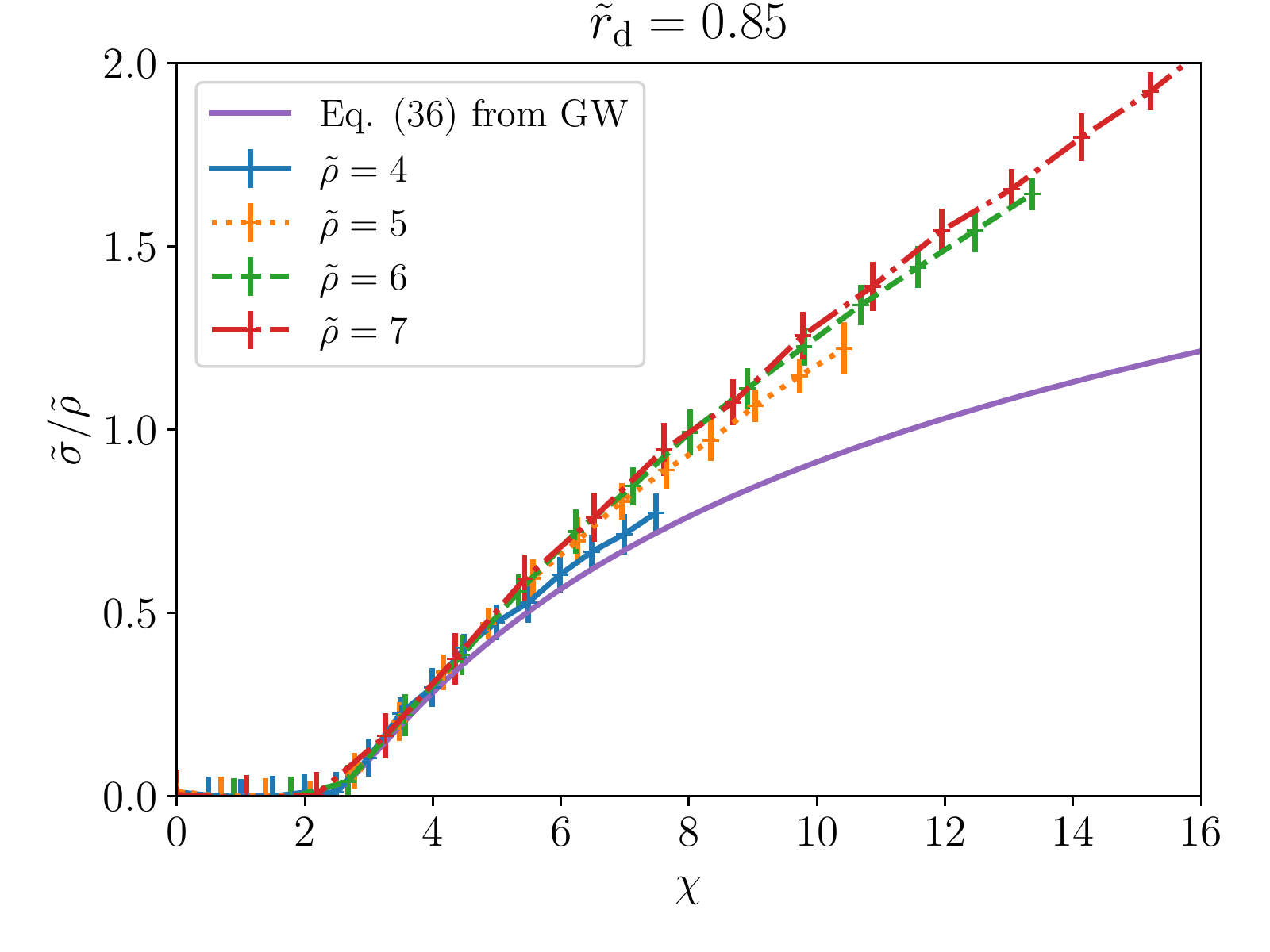}
\caption{}
\label{fig:sigma_rd_85}
\end{subfigure}
\caption{Dependence of density-scaled surface tension $\tilde\sigma/\tilde\rho$ on the $\chi$-parameter for (a) $\trd=0.75$ and (b) $\trd=0.85$ with error bars.}
%\label{fig:sigma_mdpd}
\end{figure*}

\subsection{Mixing in MDPD}
\label{sec:mdpd_surf_ten}
Having provided a general liquid parametrisation protocol for MDPD and derived the interaction parameters and densities of coarse-grained water, we now turn to the mixing properties of liquids. In simulating binary mixtures, we keep the parameter $\tilde B$ constant across liquid species, as required by the no-go theorem derived by Warren.\citep{Warren_PRE_2013} Thus, phase separation can be controlled only by varying $\Delta\tilde A$.

In the context of standard DPD, mixing was related to the Flory-Huggins theory.\cite{Groot_JCP_1997} In order to bridge the experiments to mesoscale simulation, the Flory-Huggins $\chi$-parameter, which can be computed \emph{a priori} for a given mixture from Hildebrand solubilities $\delta$ via eq.~\eqref{eq:chi} or through more sophisticated Monte Carlo sampling,\cite{Fan_MA_1992} was related to the excess repulsion $\Delta\tilde A$. 

Denoting $\chi=\nu\Delta\tilde A$, $\nu=0.286$ in standard DPD at $\tilde\rho=3$ and 0.689 at $\tilde\rho=5$.\citep{Groot_JCP_1997} In the context of MDPD, Jamali \emph{et al.} derived three values of $\nu$ at three different densities, considering positive values of $\tilde A$ only and hence describing a purely repulsive liquid (eqs.~(19)--(21) in their paper).\cite{Jamali_JCP_2015} Since density in MDPD is not decided \emph{a priori} but arises by choosing the liquid and the specific CG degree, we need to understand the general dependence of $\nu$ on density. These three points obtained by Jamali \emph{et al.} can be fitted by a line:
\begin{equation}
\nu(\tilde\rho) = -0.259 + 0.196\tilde\rho.
\end{equation}

Here, we derive how $\nu$ depends not only on density but also many-body cutoff $\trd$ for negative values of $\tilde A$. Following the protocol presented by GW (Section V and Fig.~7), we set up a simulation cell with dimensions $20\times8\times8$, varied excess repulsion $\Delta\tilde A$ between 0 and 15 and measured the $\chi$-parameter from the phase-separated density profiles via:
\begin{equation}
\chi = \frac{\ln[(1-\tilde\rho_{\rm A})/\tilde\rho_{\rm A}]}{1-2\tilde\rho_{\rm A}},
\end{equation}
where $\tilde\rho_{\rm A}$ is the density of component A (for illustration, see Fig.~6 in GW\cite{Groot_JCP_1997}). Consequently, we fitted this dependence of $\chi$ on $\Delta\tilde A$ by a line. Fig.~\ref{fig:chi_da} shows that the region of linear dependence is limited to the values of $\chi$ between about 2 and 6 and also depends on the density, which should be carefully taken into consideration in simulating binary mixtures.

Exploring four different densities, we obtained a linear dependence of $\nu$ on density similar to Jamali \emph{et al.}:
\begin{equation}
\nu = -(0.233\pm0.019) + (0.188\pm0.003)\tilde\rho,
\end{equation}
for $\trd=0.75$, and:
\begin{equation}
\nu = -(0.285\pm0.019) + (0.196\pm0.003)\tilde\rho,
\end{equation}
for $\trd=0.85$. The influence of $\trd$ on $\nu$ is relatively small and for practical purposes can be neglected.

\begin{figure*}[t!]
\centering
\begin{subfigure}[b]{0.45\textwidth}
\includegraphics[width=\textwidth]{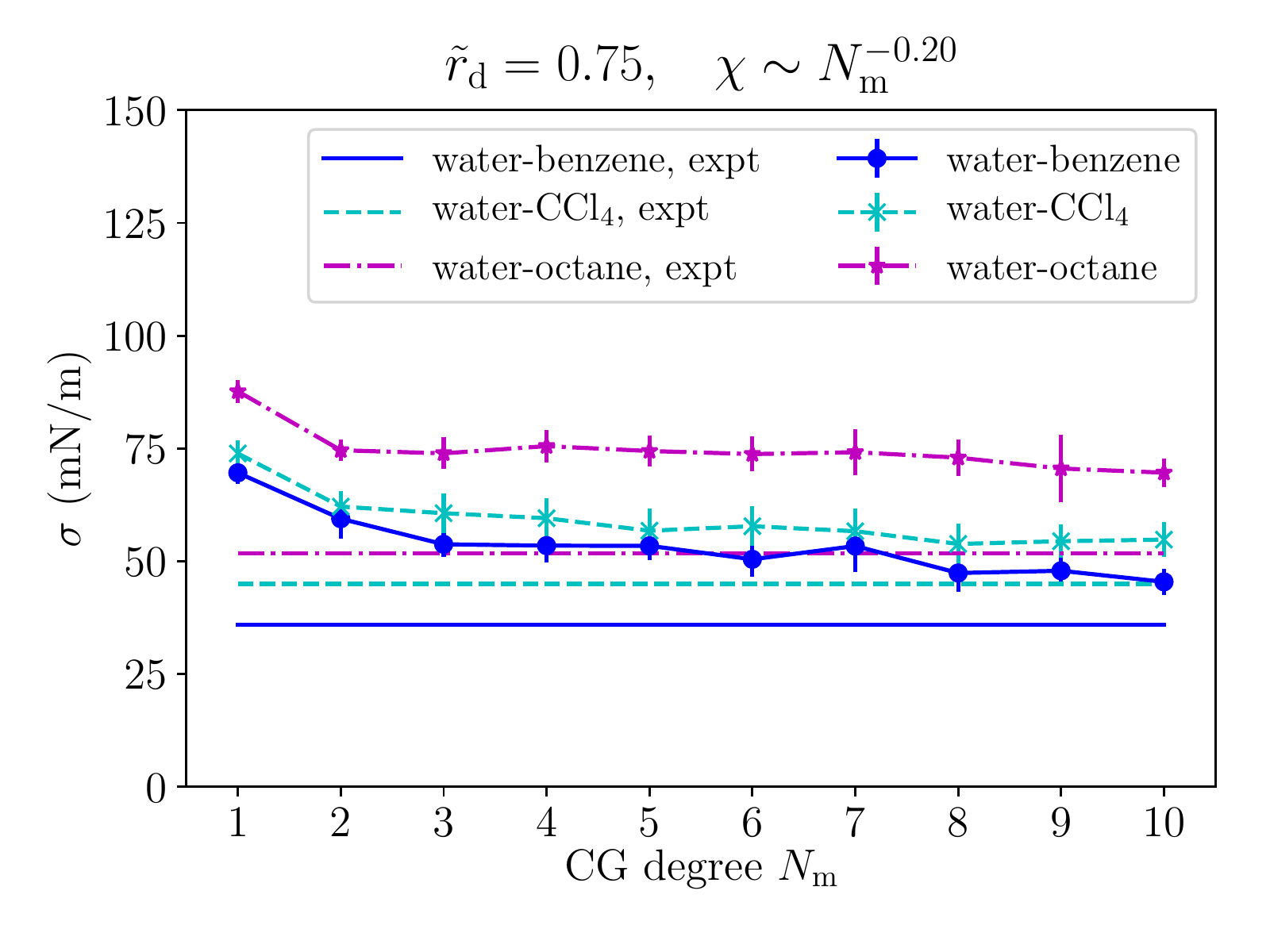}
\caption{}
\label{fig:sigma_nm_mdpd_75}
\end{subfigure}
~%\hspace{-5mm}
\begin{subfigure}[b]{0.45\textwidth}
\includegraphics[width=\textwidth]{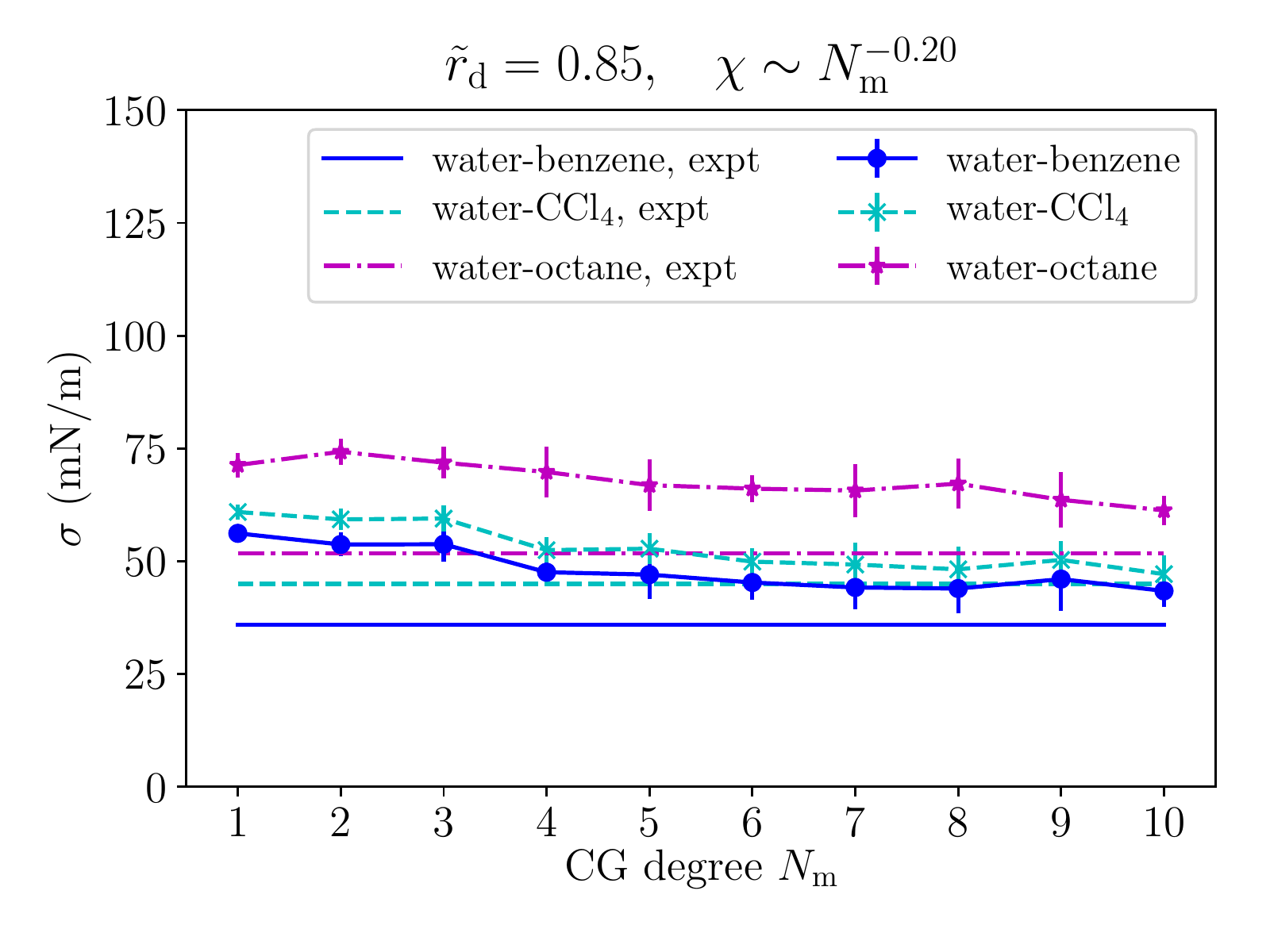}
\caption{}
\label{fig:sigma_nm_mdpd_85}
\end{subfigure}
\caption{Surface tension predicted from a MDPD simulation and compared with experiment for three solvent mixtures for (a) $\trd=0.75$ and (b) $\trd=0.85$. The scaling of the $\chi$-parameter with CG degree aims to keep real surface tensions scale invariant.}
\label{fig:sigma_nm_mdpd}
\end{figure*}

\begin{figure*}
\centering
\begin{subfigure}[b]{0.45\textwidth}
\includegraphics[width=1\textwidth]{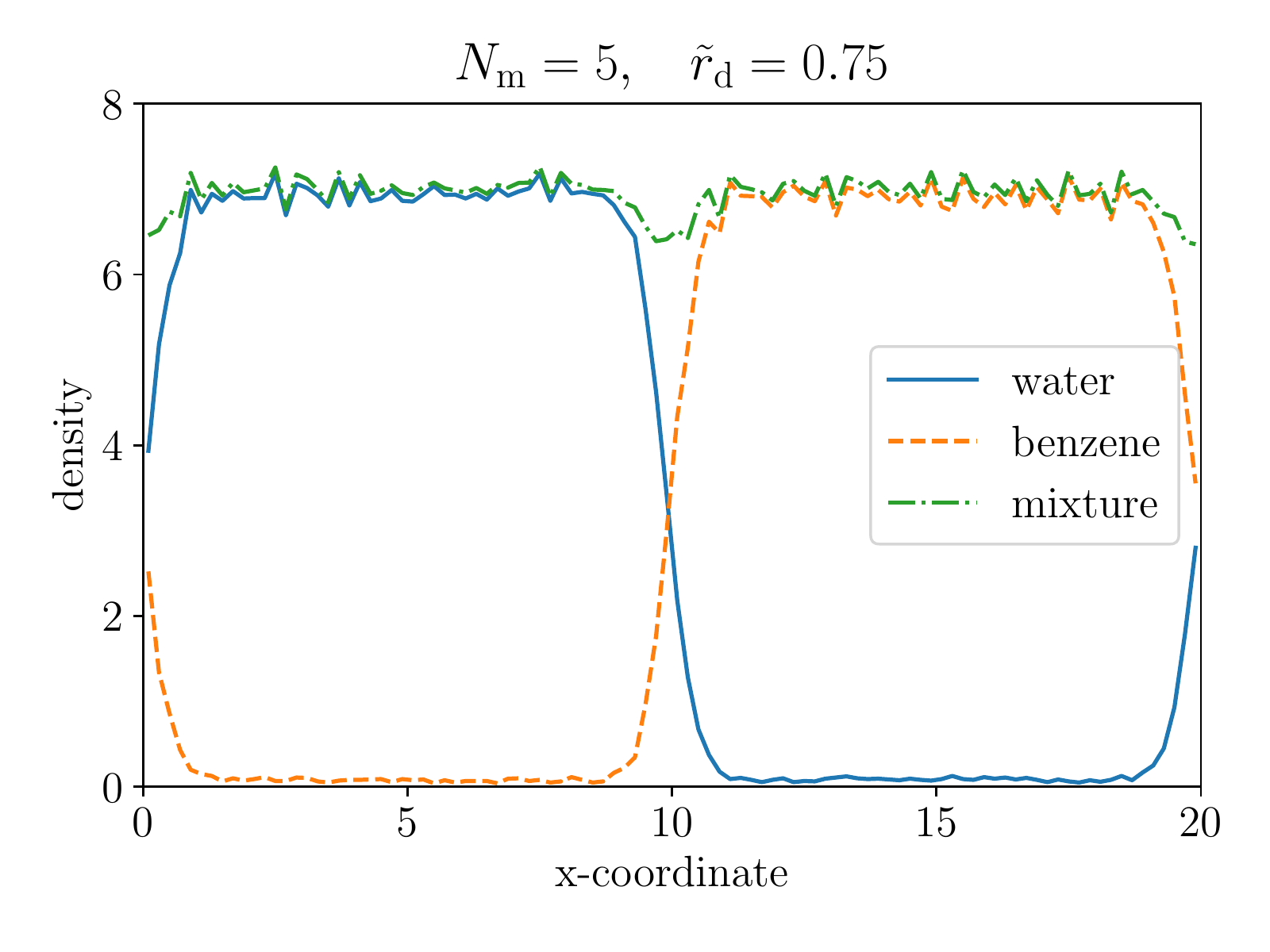}
\caption{}
\end{subfigure}
\hspace{-3mm}
\begin{subfigure}[b]{0.45\textwidth}
\includegraphics[width=1\textwidth]{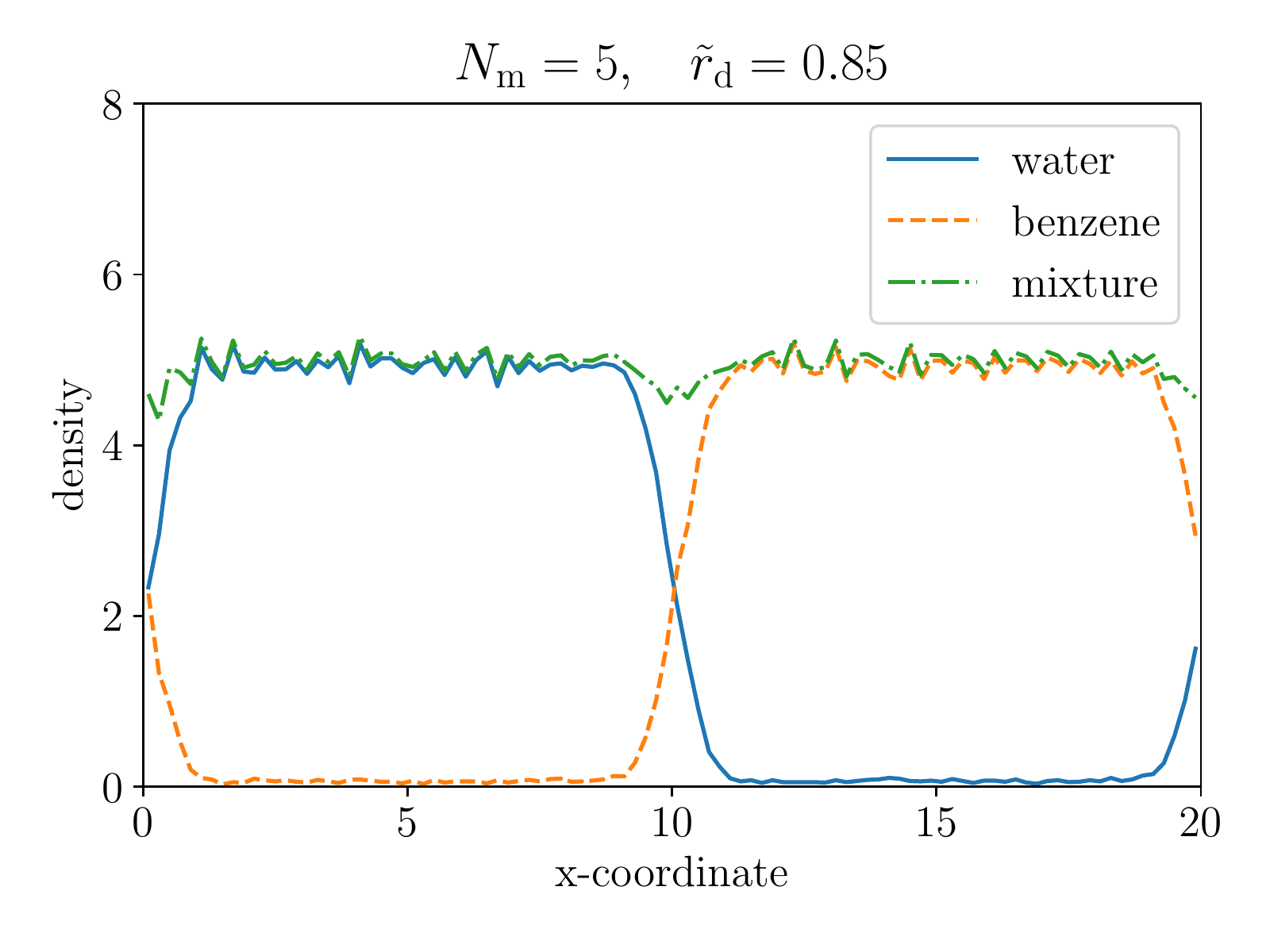}
\caption{}
\end{subfigure}
\caption{Density profiles of equilibrated many-body DPD mixture of water and benzene at CG degree $\Nm=5$ for many-body cutoffs (a) $\trd=0.75$ and (b) $\trd=0.85$.}
\label{fig:wb_profiles}
\end{figure*}

\subsection{Surface tension}
Having determined the dependence of the $\chi$-parameter on excess repulsion $\Delta\tilde A$, we now turn our attention to surface tension, a key quantitative descriptor of behaviour of a binary mixture. 

Firstly, we verify how surface tension varies on $\chi$-parameter. We note that Jamali \emph{et al.} have also computed this dependence (Fig. 12b in\cite{Jamali_JCP_2015}) but did not provide a functional form. We decided to revisit their results due to a different choice of interaction parameters $\tilde A>0$ by these authors. As in Section~\ref{sec:dpd_scaling}, we used the pressure tensor components for surface tension calculation via eq.~\eqref{eq:sigma}.

Figs.~\ref{fig:sigma_rd_75} and~\ref{fig:sigma_rd_85} show the surface tension vs $\chi$-parameter for $\trd=0.75$ and 0.85, respectively. We do not observe the collapse of the ratio $\tilde\sigma/\tilde\rho$ onto one curve, as GW claimed, beyond $\chi>5$, as there still remains a small density dependence. Furthermore, our absolute values of the surface tension are lower by about a factor of three from the values obtained by Jamali \emph{et al.} (Fig.12b), but in agreement with Fig.~1b from Yong.\cite{Yong_Polymers_2016}

GW suggested a fitting form $\tilde\sigma=\mu_1 \chi^{\mu_2}(1-\mu_3/\chi)^{3/2}.$ In order to find a universal scaling where all the surface tension curves collapse onto one, we relaxed this form via coefficient $\xi$:
\begin{equation}
\tilde\sigma/\tilde\rho^\xi = \mu_i \chi^{\mu_2}(1-\mu_3/\chi)^{3/2}.
\label{eq:sigma_gw}
\end{equation}
Searching for $\xi$ that minimises the standard deviation on $\mu_i$, which is a signature of universal scaling, we found that the best fit is provided by $\xi=1.38$ for $\trd=0.75$ and 1.24 for $\trd=0.85$.

As in the case of standard DPD, to enable reliable simulations of real mixtures at various scales, we need to derive the scaling of the $\chi$-parameter with CG degree in order to keep surface tension in real units scale invariant. Fitting for $\mu_i$ in eq.~\eqref{eq:sigma_gw} and computing surface tension for the three mixtures considered by Maiti \emph{et al.}\cite{Maiti_JCP_2004}, namely: water--benzene, water--CCl$_4$ and water--octane, at CG degrees 1--10, we found that $\chi\sim\Nm^{-0.2}$ yields the smallest RMSE with respect to experimental values in Table~\ref{tbl:liquids}, an exponent similar to $-0.22$ for standard DPD.

Finally, to verify the predictive capability of MDPD, we computed via simulation the surface tensions of mixtures for a range of CG degrees and the two viable many-body cutoffs, 0.75 and 0.85. We remark that the $\chi$-parameters computed by eq.~\eqref{eq:chi} of these mixtures are all on the high end of the range of validity in Fig.~\ref{fig:chi_da} at low CG degrees. Setting the simulation cell $20\times10\times10$ and timestep $\Delta\tilde t =0.02$, we simulated in DL\_MESO version 2.6\cite{DLMS} for 150k time steps, using first 50k for equilibration and collecting in 10k increments the pressure tensor components for averaging. The interaction parameters $\tilde B_{ij}$ were the same for all pairs of species due to Warren's no-go theorem, and $\tilde A_{ij}$ were different only for unlike species:
\begin{align}
\tilde A_{ij} &= \tilde A + \nu(\tilde\rho)\chi_{ij},\\
\tilde B_{ij} &= \tilde B,
\end{align}
where $\tilde A,\tilde B$ were taken from Tables~\ref{tbl:sol_75} or~\ref{tbl:sol_85} for appropriate CG degree.

For $\trd=0.75$, the results on Fig.~\ref{fig:sigma_nm_mdpd_75} show a satisfactory albeit not perfect agreement, only apart from $\Nm=1$ and 2, where the deviation is more significant. At these low CG degrees, the densities are very high and already out of the range of validity of the density fit,\cite{Vanya_PRE_2018} resulting in incorrect liquid behaviour. Increasing the many-body cutoff to $\trd=0.85$, Fig.~\ref{fig:sigma_nm_mdpd_85} shows good agreement of up to 10\% in case of water--CCl$_4$. Considering that due to lower density $\tilde\rho$ the simulations took about a third of the time required by the configurations employing $\trd=0.75$, this setting is suitable for water-solvent simulations. Illustrative density profiles of water and benzene at $\Nm=5$ are shown in Fig.~\ref{fig:wb_profiles}.

Finally, we note that treating water and other solvents with the same set of default interaction parameters $(A_{ii}, B_{ii})$ is sufficient if the simulation cell is filled with liquid phase only, as is the case of our current simulations. However, to simulate liquid-vapour coexistence it would be ideal if the two solvents had their own set of default parameters derived from their respective compressibilities and surface tensions. At present, this is a challenge for MDPD due to the no-go theorem\cite{Warren_PRE_2013} preventing different values of $B_{ij}$.

\section{Conclusions}
In this work, we explored the freedom in tuning the force field of both standard and many-body dissipative particle dynamics. We reviewed the derivation of the temperature-dependence of the interaction parameter, first proposed by Groot and Warren.\cite{Groot_JCP_1997} Consequently, we theoretically revisited the scaling of the simulation variables and elucidated the role of the coarse-graining degree, an important ingredient of a mesoscale simulation. We derived the scaling of the friction and interaction parameters so that the experimental observables emerging from the simulation would remain invariant with respect to the coarse graining.

For the many-body DPD, we explored a range of the many-body cutoffs and derived the interaction parameters simulating water at correct surface tension and compressibility while preserving simulation efficiency by minimising the number of particles in a simulation cell. Building on this, we derived the scaling of the Flory-Huggins $\chi$-parameter, which controls the mixing of liquids, on excess repulsion as well as coarse-graining degree. Our findings will enable the application of the many-body DPD to more complex soft matter systems including pores, liquid/solid or liquid/vapour interfaces on the length scales of 10--100 nm, such as, for example, polymer electrolyte membranes, and raise the predictive accuracy \emph{vis \`a vis} experimental data.

\section{Acknowledgments}
P.V. and J.A.E. acknowledge support of EPSRC and Johnson Matthey. P.V. acknowledges financial support by	 Sir Colin Corness Bursary.

\bibliography{ref.bib}
\end{document}